\documentclass{article}
\usepackage[margin=1in]{geometry}
\usepackage[utf8]{inputenc}
\usepackage{import}
\usepackage{xcolor}
\usepackage{tikz}
\usepackage{amsmath}
\usepackage{amsfonts}
\usepackage{hyperref}
\usepackage{booktabs}
\usepackage{multirow}
\usepackage{longtable}
\usepackage{afterpage}
\usetikzlibrary{arrows}
\newcommand{\rev}{\textcolor{black}}

\begin{document}
\begin{center}
    \textbf{\rev{Digital Twins for the Designs of Systems: a Perspective}}
    
\end{center}

\label{authors}

\noindent\textbf{Anton van Beek}\\
School of Mechanical and Materials Engineering \\
University College Dublin\\ D04 V1W8, Dublin\\
email: anton.vanbeek@ucd.ie\\

\noindent\textbf{Vispi Karkaria}\\
Department of Mechanical Engineering \\
Northwestern University\\ 60208, Evanston, Illinois\\
email: vispikarkaria2026@u.northwestern.edu \\

\noindent\textbf{Wei Chen\footnote{Corresponding author} }\\
Department of Mechanical Engineering \\
Northwestern University\\ 60208, Evanston, Illinois\\
email: weichen@northwestern.edu

\clearpage

\section*{Abstract}
\label{abstract}
The design and operation of systems are conventionally viewed as a sequential decision-making process that is informed by data from physical experiments and simulations. However, the integration of these high-dimensional and heterogeneous data sources requires the consideration of the impact of a decision on a system’s remaining life cycle. Consequently, this introduces a degree of complexity that in most cases can only be solved through an integrated decision-making approach. In this \rev{perspective} paper, we use the digital twin concept to formulate an integrated perspective for the design of systems. Specifically, we show how the digital twin concept enables the integration of system design decisions and operational decisions during each stage of a system's life cycle. This perspective has two advantages: (i) improved system performance as more effective decisions can be made, and (ii) improved data efficiency as it provides a framework to utilize data from multiple sources and design instances. From the formal definition, we identify a set of eight capabilities that are vital constructs to bring about the potential, as defined in this paper, that the digital twin concept holds for the design of systems. Subsequently, by comparing these capabilities with the available literature on digital twins, we identify a set of research questions and forecast what their broader impact might be. \rev{By conceptualizing the potential that the digital twin concept holds for the design of systems, we hope to contribute to the convergence of definitions, problem formulations, research gaps, and value propositions in this burgeoning field.} Addressing the research questions, associated with the digital twin-inspired formulation for the design of systems, will bring about more advanced systems that can meet some of the societies' grand challenges.

\section{Introduction and Motivation}
\label{intro}
The advent of the internet of things and advancements in computational resources has accelerated the design of systems with increasing degrees of complexity \cite{fuller2019digital}. A specific concept that has received increased attention from the scientific community is the ``digital twin''. The digital twin concept includes the perspective that a set of physics-based simulation models can be used to model the state of a system, and be combined with dynamically collected data to provide operation support \cite{tao2019digital,son2022past}. Examples of where the digital twin concept has been applied can be found in aerospace, manufacturing, and automotive engineering \cite{van2021archetypes,haag2019automated}. \rev{The digital twin concept can be viewed as a special case of a cyber-physical-social system that involves bidirectional communication between the digital and physical layers of the system \cite{Abdulmotaleb2017}. Note that this can also include bidirectional communication with the social layer.} While increasing in popularity for the operation of systems, the merit that the digital twin concept holds for the design of systems is not well-defined \cite{wright2020tell}. In this paper, we establish a digital twin-inspired framework for the design of systems for integrated decision-making of design and operation decisions.

Within the context of engineering design, the digital twin concept has been perceived as a variety of different concepts (e.g., a three-dimensional computer-aided design model, a machine learning model, or a physics-based simulation) \cite{barkanyi2021modelling}. According to the definition provided in \cite{schweigert2020conception}, the digital twin concept encompasses all these interpretations. Specifically, the digital twin concept includes the functionality that a digital system can be realized in the form of a virtual information construct that dynamically mimics the structure, context, and behavior of a physical asset through observed data. The purpose of the digital system is to inform decisions that improve the value of the physical system \cite{AIAA2020}. However, this perspective does not consider the design of the physical system, nor does it enable systematic integration of data from previous design generations.

For the digital system to provide an accurate representation of a specific physical system, it is provided with data that is dynamically collected throughout a system’s life cycle. Through real-time data analysis, the digital system can then be used to mimic the physical state and inform decisions that have improved value for a specific system (e.g., vehicles of the same model could benefit from different operation decisions as they are used by different drivers with different driving behavior under different usage contexts) \cite{rosen2015importance,sharma2020digital, soderberg2017toward}. However, the dynamic updating of the digital system places stringent demands on the required computational resources, limiting the effectiveness of the digital twin concept \cite{reuther2019survey,bergquist2001vector}. In addition, as it is desired to consider the effect of decisions on a system’s entire life cycle, a model that can predict the future state of the physical system is required. However, this introduces uncertainty, as no model is perfect \cite{biesinger2019facets}, and thus places an additional demand on the necessary computational resources to quantify this uncertainty \cite{halemane1983optimal}. However, as more data is collected over a system’s life cycle, the uncertainty of the predicted future states will decrease as the models will have improved fidelity (e.g., data-driven models improve in fidelity as they are dynamically updated with observed data) \cite{famelis2019managing}.

The prediction of the state of a system in the future has reduced fidelity for longer time horizons and could explain why much of the work on the digital twin concept focuses on the later stages of a system’s life cycle (e.g., predictive maintenance \cite{Magargle2017, barthelmey2019dynamic}, structural health monitoring \cite{tuegel2011reengineering, Seshadri2017, Wei2022}, recycling \cite{Wang2019}, and recovery \cite{Ayani2018}). However, with the increasing availability of computational resources and algorithmic advancements, progressively more digital twin-related work is being done at earlier stages of a system’s life cycle (e.g., manufacturing\cite{leng2021digital, ghanem2020probabilistic, guo2020framework,moser2020mechanistic}, supply chain management \cite{barykin2021place}, and policymaking in cyber-physical social systems \cite{karkaria2021computational, tao2019digital}). \rev{ As decisions at the early stage of a system’s life cycle have a higher impact on the
overall system performance compared to downstream decisions, it becomes critical that consideration is given to the decisions in the design stage (upper stream) of a system. }

In this paper, we introduce a digital twin-inspired perspective to the design of systems. Specifically, we show that by integrating online data collected from sensors into the objective function for decision-making we can jointly measure the relative merit of design and operating decisions. This has two primary advantages: (i) it provides an integrated framework for the design of systems by joint consideration of design and operation decisions, and (ii) it provides improved data efficiency by enabling the integration of data from previous system generations (e.g., manufactures of phones, automobiles, and airplanes that periodically release new generations of the same designs). Design and operation decisions warrant joint consideration as they influence a system's performance (e.g., the operation decisions made during the manufacturing process depend on the part geometry determined during the design phase). Moreover, the proposed framework centers around the optimization of an objective function with respect to the system design and operation decisions. Note that the introduced framework involves a design perspective that accounts for the multiple phases of a system's life cycle as opposed to the typical single-phase user-centered implementation of the digital twin concept \cite{jaensch2018digital,zhou2019digital,Wang2019}. In this paper, we provide a formal formulation for the decision-making model based on the digital twin concept. Through analysis of the objective function formulation, we identify a set of capabilities that are necessary to advance the potential that the digital twin concept holds for the design of systems. By comparing the identified capabilities with the available literature on the digital twin concept we unearth a set of research opportunities for further scientific inquiry. Continued research in the identified research areas will enable the discovery of generalizable knowledge that will bring about the design of more advanced systems.

In Section~\ref{DTandDesign}, we will introduce the digital twin-inspired perspective to the design of systems and use it to identify a set of capabilities that are vital to its realization. In Section~\ref{CurrentState}, the current state of the digital twin concept is discussed by comparing available methods with the identified capabilities. Subsequently, in Section~\ref{Future} we identify a set of research directions considering the introduced digital twin-inspired perspective and the limitations of the digital twin-inspired methods in the literature. Finally, in Section~\ref{Conlusions} we conclude this work by summarizing our contributions, reiterating the identified research directions, and forecasting their potential impact.

\section{Digital Twins and Design}
\label{DTandDesign}
In this section, we provide a formal definition of the digital twin concept and state our premises for how this can be advantageous for the design of systems. In addition, we will use this definition to identify a set of capabilities that are vital to the effective deployment of the digital twin-inspired perspective to system design. 

\subsection{Digital Twins as a Concept for Design}
While multiple definitions of the digital twin concept have been proposed in previous literature, we adopt the definition of the American Institute of Aeronautics and Astronautics \cite{AIAA2020}. Specifically, they defined the digital twin concept as,
\begin{center}
    \textit{A set of virtual information constructs that mimics the structure, context and behavior of an individual / unique physical asset, or a group of physical assets, is dynamically updated with data from its physical twin throughout its life cycle and informs decisions that realize value.}
\end{center}
We believe that this is an appropriate definition as it emphasizes that a digital twin is unique to a specific physical system, involves the acquisition of data, and the use of this data to inform decisions that realize value over the entire life cycle of a system. Moreover, this definition provides sufficient freedom to be compatible with any type of system (e.g., manufacturing, aerospace, and automotive systems), while being specific enough in terms of its fundamental constructs (i.e., dynamic data acquisition, prediction, and decision-making).

In the context of system design, the authors believe that the value proposition that the digital twin concept holds is as follows
\begin{center}
    \textit{The digital twin concept provides an integrated perspective for the design of systems to improve data efficiency and performance through the joint consideration of design and operation decisions.}
\end{center}
Specifically, the digital twin concept for the design of systems enables a systematic approximation of a system’s state/merit throughout each step in its life cycle. Consequently, such an approximation can be used to improve the value proposition of a specific system. In the context of a system’s design, the digital twin concept encapsulates the consideration of operating conditions during the design of the system. Consequently, this enables the formulation of design decisions that not only optimize the value proposition of the system at its inception but rather over its entire life cycle. In addition, the availability of a model that approximates the state of a system’s life cycle enables a multi-generation design perspective. Specifically, data collected from previous generations of a design can be used to establish the physical and digital system of the next generation with improved performance and fidelity, respectively (e.g., cars, phones, and airplanes have new design generations periodically). We focus our attention on the multi-generation design of the same systems as using data from dissimilar systems will be more complicated and of reduced utility. Nevertheless, the digital twin concept provides a perspective that facilitates the collection and utilization of data over a system's life cycle and across multi-generations.

A specific advantage that the digital twin concept holds for the design of systems is that it enables a designer to acknowledge the interdependence between design and operation decisions. For example, decisions about the disposal (i.e., operation decision) of a system depend on the materials used (i.e., design decision) in the physical components of the system. When taking a discrete-time perspective to the operation of a system, we find that for each time $t$ a decision $\textbf{u}_t\in \textbf{U}$ has to be rendered. The space of admissible operation decisions $\textbf{U}$ can, among others, include manufacturing decisions $\textbf{U}_m$, control decisions throughout the service life of the system $\textbf{U}_s$, and disposal decisions $\textbf{U}_d$ (i.e., $\textbf{U}\ = \left\{\textbf{U}_m \cup \textbf{U}_s \cup \textbf{U}_d \right\})$. However, the operational decisions are dependent on the design of the system that can be represented by a vector $\textbf{x}\in\textbf{X}\subset \mathbb{R}^{d_x}\times \mathbb{N}^{p_x}$ where $d_x$ is the dimensionality of the quantitative design variables and $p_x$ is the dimensionality of qualitative design variables (i.e., a set of mutually exclusive categorical alternatives).

The system design and operation decisions require an objective function $f:\textbf{S}_t\times\textbf{U}_t\times \textbf{X}_t\rightarrow \mathbb{R}$ that measures the relative merit of admissible decisions. Consequently, this introduces an optimization problem for the design of a single generation of a system as
\begin{align}
    \textbf{x}^*,\boldsymbol{\mu}^*& =\underset{\textbf{x}\in\textbf{X},\mu\in M}{\mathrm{argmax}} f(\textbf{s}_t,\mu_t(\textbf{s}_t),\textbf{x}), \nonumber\\
    &=\underset{\textbf{x}\in\textbf{X},\mu\in M}{\mathrm{argmax}}\sum_{t=1}^{T}r_t(\textbf{s}_t,\mu_t(\textbf{s}_t),\textbf{x}),
    \label{POMDP_approx}
\end{align}
where we used a policy function $\mu:S_t\rightarrow U$ that maps a state of the system $s_t\in S_t$ into a control decision at time $t$. In addition, we assume that we can approximate the state of the physical system through a set of data obtained from sensors placed on the physical system $\textbf{P}_t \in \mathcal{P}_t$, and a set of simulation data $\textbf{D}_t\in \mathcal{D}_t$ (i.e., a state if defined as $\textbf{s}_t\in S_t = \mathcal{P}_t\times\mathcal{D}_t\subset \mathbb{R}^{d_s}\times \mathbb{N}^{p_s}$ where $d_s$ is the dimensionality of the quantitative state variables and $p_s$ is the dimensionality of the qualitative state variables that represents a set of mutually exclusive categorical alternatives). \rev{It should be observed that how a system's state is defined is consistent with the definition of the digital twin concept as its construction requires ``virtual information constructs'' (e.g., simulations and virtual models).} Moreover, as decisions have to be made for a dynamic problem with a time horizon $T$, we require a reward function $r_t:S_t\times \textbf{U}_t \rightarrow \mathbb{R}$ that measures the relative merit of competing prospects for a discrete point in time. \rev{Note that this implies that $r_t(\cdot)$ can be used throughout the different stages of a system's service life. This can be accommodated by having a conditional operator that defines a unique reward function for each stage in a system's life cycle. For example, the reward function can be defined as
\begin{equation}
    r_t(\cdot)=\begin{cases}
    r_t^{(m)}(\cdot),& \text{if in manufacturing stage} \\ r_t^{(s)}(\cdot),& \text{if in service stage} \\ r_t^{(d)}(\cdot),& \text{if in disposal stage} 
\end{cases}
\end{equation}
where $r_t^{(m)}(\cdot), r_t^{(s)}(\cdot), r_t^{(d)}(\cdot)$ are the reward functions for the manufacturing, service, and disposal stage of a system's life cycle, respectively. Note that the integration of design and operation decisions is similar to the field of control co-design \cite{Cui2020}. The difference between control co-design and the method proposed in this paper is that control co-design involves the design of a system's control during one stage of the system's service life whereas we propose the joint consideration of control decisions across all life cycle stages. In addition, while co-design involves the design of control systems, the presented framework also enables human-computer interaction through, for example, nudging \cite{kissmer2018}. Control co-design has shown to significantly improve the value proposition of various autonomous systems \cite{Garcia2019}, and as such foreshadows the benefits that can be attained by joint consideration of design and operation decisions for the design of systems.} The final piece for the optimization of Equation~\ref{POMDP_approx} is a state transition function $\Gamma(\cdot)$ that maps the system state $\textbf{s}_t$ and operation decision $\textbf{u}_t$ into an approximation of the state at the next time step $t+1$ as $\textbf{s}_{t+1}=\Gamma_t(\textbf{s}_t,\textbf{u}_t)$ (i.e., we require a dynamic as $\Gamma_t : S_t\times U_t\rightarrow S_{t+1}$). Note that the reward function $r(\cdot)$ is not the same as the objective function $f(\cdot)$ as it only gives the value of being in a specific state $\textbf{s}_t$ at time $t$, whereas the objective function $f(\cdot)$ considers the evolution of the system over its life cycle.
 
The digital twin concept for the design of systems requires the formulation of the reward function $r_t(\cdot)$ and the state transition function $\Gamma_t(\cdot)$. For the first generation of a system's design, these models will primarily be driven by physics-based simulations. However, as more sensor data is collected throughout the life cycle of the system, we can update these models by synthesizing experimental and simulation data in a hybrid model. The fidelity and utility of these models will increase as more data is collected. The full potential of the digital twin concept would be achieved once a digital representation has been obtained that is identical to the physical system. However, from this statement, we can make two observations: (i) it is unrealistic to achieve a digital system that unlocks the full potential of the digital twin concept as this would require a state transition function $\Gamma_t(\cdot)$ that perfectly represents physical reality, and therefore (ii) when optimizing Equation~\ref{POMDP_approx} we need to account for the uncertainty associated with the model parameters, model inadequacies, and collected data. One approach to account for these sources of uncertainty is to take the expectation of the objective function in Equation~\ref{POMDP_approx} as
\begin{align}
    \textbf{x}^*,\boldsymbol{\mu}^* =
    \underset{\textbf{x}\in\textbf{X},\mu\in M}{\mathrm{argmax}}\mathbb{E}\left[\sum_{t=1}^{T}r_t(\textbf{s}_t,\mu_t(\textbf{s}_t),\textbf{x})\right].
    \label{POMDP_approx2}
\end{align}
While taking the expectation of the objective function enables a designer to account for the central tendency of the reward, it neglects the magnitude of the variability in the predicted response. Consequently, alternative strategies include robust optimization \cite{Tsui1999}, reliability-based design \cite{Youn2004}, and utility-based decision making \cite{Hazelrigg1998}.

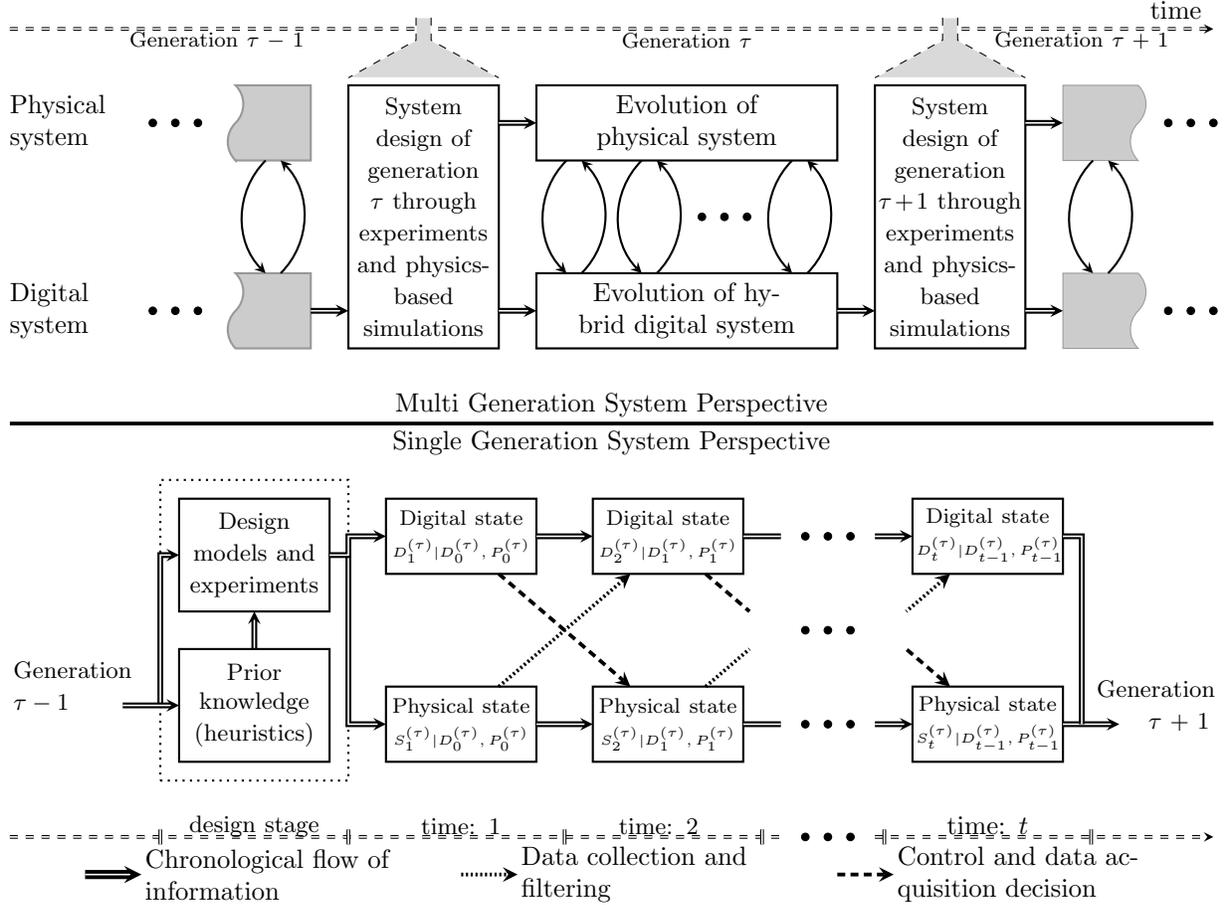
\begin{figure}[t]
\begin{center}
\setlength{\unitlength}{0.012500in}%
\begin{tikzpicture}

\node[align=left,text width=20mm,thick] at (10mm,105mm) {Physical\\system};
\node[align=left,text width=20mm,thick] at (10mm,80mm) {Digital\\system};
\filldraw[fill=black!50!white,draw=black, opacity=0.4,thick] (30mm,75mm) --++ (10mm,0mm)  --++ (0mm,10mm) --++ (-10mm,0mm) arc (45:-45:3.535533mm) --++ (0mm,0mm) arc (135:225:3.5355339mm) -- cycle;
\filldraw[fill=black!50!white,draw=black, opacity=0.4,thick] (30mm,100mm) --++ (10mm,0mm)  --++ (0mm,10mm) --++ (-10mm,0mm) arc (45:-45:3.535533mm) --++ (0mm,0mm) arc (135:225:3.5355339mm) -- cycle;
\foreach \x in {0,...,2}{\filldraw[black] (19mm+\x*3mm,80mm) circle (0.5mm);}
\foreach \x in {0,...,2}{\filldraw[black] (19mm+\x*3mm,105mm) circle (0.5mm);}

\draw[fill=white, draw=black,thick] (45mm,75mm) --++ (20mm,0mm)  --++ (0mm,35mm) --++ (-20mm,0mm) -- cycle;
\draw[fill=white, draw=black,thick] (115mm,75mm) --++ (20mm,0mm)  --++ (0mm,35mm) --++ (-20mm,0mm) -- cycle;
\draw[fill=white, draw=black,thick] (70mm,75mm) --++ (40mm,0mm)  --++ (0mm,10mm) --++ (-40mm,0mm) -- cycle;
\draw[fill=white, draw=black,thick] (70mm,100mm) --++ (40mm,0mm)  --++ (0mm,10mm) --++ (-40mm,0mm) -- cycle;

\filldraw[fill=black!50!white,draw=black, opacity=0.4] (150mm,75mm) --++ (-10mm,0mm)  --++ (0mm,10mm) --++ (10mm,0mm) arc (45:-45:3.535533mm) --++ (0mm,0mm) arc (135:225:3.5355339mm) -- cycle;
\filldraw[fill=black!50!white,draw=black, opacity=0.4] (150mm,100mm) --++ (-10mm,0mm)  --++ (0mm,10mm) --++ (10mm,0mm) arc (45:-45:3.535533mm) --++ (0mm,0mm) arc (135:225:3.5355339mm) -- cycle;
\foreach \x in {0,...,2}{\filldraw[black] (154mm+\x*3mm,80mm) circle (0.5mm);}
\foreach \x in {0,...,2}{\filldraw[black] (154mm+\x*3mm,105mm) circle (0.5mm);}

\draw[>=stealth,double, ->,thick] (40mm,80mm) --++ (5mm,0mm);
\draw[>=stealth,double, ->,thick] (65mm,80mm) --++ (5mm,0mm);
\draw[>=stealth,double, ->,thick] (65mm,105mm) --++ (5mm,0mm);
\draw[>=stealth,double, ->,thick] (110mm,80mm) --++ (5mm,0mm);
\draw[>=stealth,double, ->,thick] (135mm,80mm) --++ (5mm,0mm);
\draw[>=stealth,double, ->,thick] (135mm,105mm) --++ (5mm,0mm);

\draw[>=stealth, ->,thick] (74mm,100mm) arc (135:225:10.6066mm);
\draw[>=stealth, ->,thick] (76mm,85mm) arc (-45:45:10.6066mm);
\draw[>=stealth, ->,thick] (84mm,100mm) arc (135:225:10.6066mm);
\draw[>=stealth, ->,thick] (86mm,85mm) arc (-45:45:10.6066mm);
\draw[>=stealth, ->,thick] (104mm,100mm) arc (135:225:10.6066mm);
\draw[>=stealth, ->,thick] (106mm,85mm) arc (-45:45:10.6066mm);

\draw[>=stealth, ->,thick] (144mm,100mm) arc (135:225:10.6066mm);
\draw[>=stealth, ->,thick] (146mm,85mm) arc (-45:45:10.6066mm);
\draw[>=stealth, ->,thick] (34mm,100mm) arc (135:225:10.6066mm);
\draw[>=stealth, ->,thick] (36mm,85mm) arc (-45:45:10.6066mm);

\foreach \x in {0,...,2}{\filldraw[black] (92mm+\x*3mm,92.5mm) circle (0.5mm);}

\node[align=center,text width=19mm,thick] at (55mm,92.5mm) {\small System design of generation $\tau$ through experiments and physics-based simulations};
\node[align=center,text width=19mm,thick] at (125mm,92.5mm) {\small System design of generation $\tau+1$ through experiments and physics-based simulations};
\node[align=center,text width=39mm,thick] at (90mm,105mm) {Evolution of physical system};
\node[align=center,text width=39mm,thick] at (90mm,80mm) {Evolution of hybrid digital system};

\draw[>=stealth, ->, double, dashed] (0mm,117.5mm) --++ (160mm,0mm);
\node[align=center,text width=39mm,thick] at (155mm,120mm) {time};
\node[align=center,text width=39mm,thick] at (27.5mm,116mm) {\footnotesize Generation $\tau-1$};
\node[align=center,text width=39mm,thick] at (90mm,116mm) {\footnotesize Generation $\tau$};
\node[align=center,text width=39mm,thick] at (142.5mm,116mm) {\footnotesize Generation $\tau+1$};

\draw[fill=white!85!black, draw=white] (54mm,119mm) --++ (0mm,-3mm)  --++ (-9mm,-5mm) --++ (20mm,0mm) --++ (-9mm,5mm) --++ (0mm,3mm) -- cycle;
\draw[black=white,dashed] (54mm,119mm) --++ (0mm,-3mm)  --++ (-9mm,-5mm);
\draw[black=white,dashed] (56mm,119mm) --++ (0mm,-3mm)  --++ (9mm,-5mm);

\draw[fill=white!85!black, draw=white] (124mm,119mm) --++ (0mm,-3mm)  --++ (-9mm,-5mm) --++ (20mm,0mm) --++ (-9mm,5mm) --++ (0mm,3mm) -- cycle;
\draw[black=white,dashed] (124mm,119mm) --++ (0mm,-3mm)  --++ (-9mm,-5mm);
\draw[black=white,dashed] (126mm,119mm) --++ (0mm,-3mm)  --++ (9mm,-5mm);

\draw[line width=0.5mm] (0mm,65mm) --++ (160mm,0mm);
\node[align=center,text width=60mm,thick] at (80mm,67.5mm) {Multi Generation System Perspective};
\node[align=center,text width=60mm,thick] at (80mm,62.5mm) {Single Generation System Perspective};


\draw[fill=white, draw=black,thick,dotted] (20mm,17.5mm) --++ (25mm,0mm)  --++ (0mm,40mm) --++ (-25mm,0mm) -- cycle;
\draw[>=stealth,double, ->,thick] (20mm,27.5mm) --++ (0mm,20mm) --++ (2.5mm,0mm);
\draw[>=stealth,double, ->,thick] (15mm,27.5mm) --++ (7.5mm,0mm);
\draw[>=stealth,double, ->,thick] (32.5mm,35mm) --++ (0mm,5mm);
\draw[fill=white, draw=black,thick] (22.5mm,20mm) --++ (20mm,0mm)  --++ (0mm,15mm) --++ (-20mm,0mm) -- cycle;
\draw[fill=white, draw=black,thick] (22.5mm,40mm) --++ (20mm,0mm)  --++ (0mm,15mm) --++ (-20mm,0mm) -- cycle;

\draw[>=stealth,double,thick,<-] (50mm,25mm)--++ (-5mm,0mm) --++ (0mm,22.5mm);
\draw[>=stealth,double, ->,thick] (42.5mm,47.5mm) --++ (2.5mm,0mm) --++ (0mm,2.5mm) --++ (5mm,0mm);

\draw[fill=white, draw=black,thick] (50mm,20mm) --++ (20mm,0mm)  --++ (0mm,10mm) --++ (-20mm,0mm) -- cycle;
\draw[fill=white, draw=black,thick] (50mm,45mm) --++ (20mm,0mm)  --++ (0mm,10mm) --++ (-20mm,0mm) -- cycle;
\draw[fill=white, draw=black,thick] (77.5mm,20mm) --++ (20mm,0mm)  --++ (0mm,10mm) --++ (-20mm,0mm) -- cycle;
\draw[fill=white, draw=black,thick] (77.5mm,45mm) --++ (20mm,0mm)  --++ (0mm,10mm) --++ (-20mm,0mm) -- cycle;
\draw[fill=white, draw=black,thick] (120mm,45mm) --++ (20mm,0mm)  --++ (0mm,10mm) --++ (-20mm,0mm) -- cycle;
\draw[fill=white, draw=black,thick] (120mm,20mm) --++ (20mm,0mm)  --++ (0mm,10mm) --++ (-20mm,0mm) -- cycle;

\draw[>=stealth,double,thick,->] (140mm,25mm)--++ (7.5mm,0mm);
\draw[>=stealth,double,thick] (140mm,50mm) --++ (2.5mm,0mm) --++ (0mm,-25mm);

\draw[>=stealth,double,thick,->] (70mm,25mm)--++ (7.5mm,0mm);
\draw[>=stealth,double,thick,->] (70mm,50mm)--++ (7.5mm,0mm);
\draw[>=stealth,double,thick] (97.5mm,25mm)--++ (5mm,0mm);
\draw[>=stealth,double,thick] (97.5mm,50mm)--++ (5mm,0mm);
\draw[>=stealth,double,thick,->] (115mm,25mm)--++ (5mm,0mm);
\draw[>=stealth,double,thick,->] (115mm,50mm)--++ (5mm,0mm);
\foreach \x in {0,...,2}{\filldraw[black] (105.75mm+\x*3mm,25mm) circle (0.5mm);}
\foreach \x in {0,...,2}{\filldraw[black] (105.75mm+\x*3mm,50mm) circle (0.5mm);}
\foreach \x in {0,...,2}{\filldraw[black] (105.75mm+\x*3mm,37.5mm) circle (0.5mm);}
\foreach \x in {0,...,2}{\filldraw[black] (105.75mm+\x*3mm,10mm) circle (0.5mm);}

\node[align=left,text width=15mm,thick] at (8mm,30mm) {\small Generation\\$\tau-1$};
\node[align=right,text width=15mm,thick] at (152mm,27.5mm) {\small Generation\\$\tau+1$};

\draw[>=stealth,thick,->,densely dotted,line width=0.5mm] (65mm,30mm)--++ (17.5mm,15mm);
\draw[>=stealth,thick,->,densely dashed,line width=0.5mm] (65mm,45mm)--++ (17.5mm,-15mm);

\draw[>=stealth,thick,densely dotted,line width=0.5mm] (92.5mm,30mm)--++ (17.5mm/3,15mm/3);
\draw[>=stealth,thick,densely dashed,line width=0.5mm] (92.5mm,45mm)--++ (17.5mm/3,-15mm/3);

\draw[>=stealth,thick,densely dashed,line width=0.5mm,<-] (125mm,30mm)--++ (-17.5mm/3,15mm/3);
\draw[>=stealth,thick,densely dotted,line width=0.5mm,<-] (125mm,45mm)--++ (-17.5mm/3,-15mm/3);

\draw[>=stealth,thick,double,line width=0.5mm,->] (10mm,5mm)--++ (7.5mm,0mm);
\draw[>=stealth,thick,densely dotted,line width=0.5mm,->] (60mm,5mm)--++ (7.5mm,0mm);
\draw[>=stealth,thick,densely dashed,line width=0.5mm,->] (110mm,5mm)--++ (7.5mm,0mm);
\node[align=left,text width=35mm,thick] at (35.5mm,5mm) {Chronological flow of information};
\node[align=left,text width=35mm,thick] at (85.5mm,5mm) {Data collection and filtering};
\node[align=left,text width=35mm,thick] at (135.5mm,5mm) {Control and data acquisition decision};

\node[align=center,text width=19mm,thick] at (32.5mm,27.5mm) {\small Prior knowledge \\(heuristics)};
\node[align=center,text width=19mm,thick] at (32.5mm,47.5mm) {\small Design models and experiments};

\node[align=center,text width=19mm,thick] at (60mm,50mm) {\footnotesize Digital state\\ \tiny $D_1^{(\tau)}|D_0^{(\tau)},P_0^{(\tau)}$};
\node[align=center,text width=19mm,thick] at (87.5mm,50mm) {\footnotesize Digital state\\ \tiny $D_2^{(\tau)}|D_1^{(\tau)},P_1^{(\tau)}$};
\node[align=center,text width=19mm,thick] at (130mm,50mm) {\footnotesize Digital state\\ \tiny $D_t^{(\tau)}|D_{t-1}^{(\tau)},P_{t-1}^{(\tau)}$};

\node[align=center,text width=19mm,thick] at (60mm,25mm) {\footnotesize Physical state\\ \tiny $S_1^{(\tau)}|D_0^{(\tau)},P_0^{(\tau)}$};
\node[align=center,text width=19mm,thick] at (87.5mm,25mm) {\footnotesize Physical state\\ \tiny $S_2^{(\tau)}|D_1^{(\tau)},P_1^{(\tau)}$};
\node[align=center,text width=19mm,thick] at (130mm,25mm) {\footnotesize Physical state\\ \tiny $S_t^{(\tau)}|D_{t-1}^{(\tau)},P_{t-1}^{(\tau)}$};

\draw[>=stealth, double, dashed] (0mm,10mm) --++ (102.5mm,0mm);
\draw[>=stealth, ->, double, dashed] (115mm,10mm) --++ (45mm,0mm);

\draw[>=stealth, double] (20mm,9mm) --++ (0mm,2mm);
\draw[>=stealth, double] (45mm,9mm) --++ (0mm,2mm);
\draw[>=stealth, double] (73.75mm,9mm) --++ (0mm,2mm);
\draw[>=stealth, double] (100mm,9mm) --++ (0mm,2mm);
\draw[>=stealth, double] (116.25mm,9mm) --++ (0mm,2mm);
\draw[>=stealth, double] (143.75mm,9mm) --++ (0mm,2mm);

\node[align=center,text width=39mm,thick] at (130mm,11.5mm) {time: $t$};
\node[align=center,text width=39mm,thick] at (32.5mm,11.5mm) {\small design stage};
\node[align=center,text width=30mm,thick] at (60mm,11.5mm) {\small time: $1$};
\node[align=center,text width=30mm,thick] at (86.25mm,11.5mm) {\small time: $2$};

\end{tikzpicture}
\end{center}
\caption{\textbf{A digital twin perspective to the multi-generational design of systems.} The digital twin perspective to design enables the integration of design and operation decisions as shown in the bottom half of the figure. In addition, the consideration of the entire life cycle enables the use of data over multiple generations of a system's design and is visualized in the top half of the figure.}
\label{IntroFig} 
\end{figure}

The digital twin concept for the design of systems has been visualized in Figure~\ref{IntroFig}, in the above section we show the multi-generation system design perspective, and in the bottom section, we highlight the constructs associated with a single generation of a system. In the multi-generation setting, we show that each generation starts with making the design decisions $\textbf{x}^*$ and a generalized control policy $\boldsymbol{\mu}^*$ that defines the initial state of the system $s_1$. With the generalized control policy, a distinction is made between the autonomous control decisions (e.g., the tire pressure of a vehicle) and the decisions made by the user (e.g., vehicle velocity). Specifically, during the design of a system, we can only optimize Equation~\ref{POMDP_approx2} with respect to design decisions and the autonomous operation decision, while the usage behavior can only be accounted for as a source of uncertainty. Moreover, we note that the design decisions are fixed over the life cycle of a specific system, but that the control policy for autonomous operation decisions can be updated at each time step. These decisions are made through the use of physics-based simulations and experiments. While at the first stage (i.e., $\tau = 1$) we do not yet have data collected for the evolution of generation $\tau\in\left\{1,\ldots,\tau_{max} \right\}$ over their life cycle, the designer can rely on data from previous generations or expert knowledge if $\tau > 1$. As these models will not be able to accurately represent the physical system, a designer is required to properly account for the uncertainty in their predictions that can come from uncertain parameters, model uncertainty, model bias, and noisy data.

\rev{It should be noted that the formulation presented in Equation~\ref{POMDP_approx2} describes a general class of problems that are known as Markov decision processes. This formulation is also encountered in dynamic programming, system dynamics, and reinforcement learning. Due to the generalizable formulation of the digital-twin-inspired system design framework, we can benefit from a plethora of available theories and methods.}

Once the design decisions have been made, the system goes through subsequent steps in its life cycle during which more data is collected to infer the state of the system as represented in the bottom half of Figure~\ref{IntroFig}. The digital state of a system updates itself with the online data collected from the physical system, this has two important benefits:
\begin{enumerate}
    \item The digital state updates itself recursively to represent the state of a physical system. in other words, the digital state parameter $D_t^{\tau}$ mimics the state of the system and can include parts of the physical system that are not directly observed through sensor data. 
    \item The collected data provides a comparison between the realized state of the system and the predicted state by the dynamic $\Gamma$. By comparing the predicted state to the realized state we can enhance the physics-based model with the observed data through multi-modal data fusion (i.e., improve predictive fidelity). 
\end{enumerate}
As data is being transferred from the physical system to the digital system it is typical to have a filtering step that changes the format of the collected data to be compatible with the digital system. Conversely, the digital system can be used to send a control signal to the physical system to improve operational performance or collect data that can lead to improved future performance (i.e., Equation~\ref{POMDP_approx} is optimized with respect to the control policy). For example, it could be imagined that the hybrid model has high prediction accuracy for the current mode of control, whereas its fidelity is low in an untested area of the control space. Consequently, it could be beneficial to try an untested control policy to learn if a better control policy can be identified. This is a problem that has been well studied by the optimization community in the context of Bayesian optimization \cite{Beek2021}.

\begin{figure}
\begin{center}
\includegraphics[width = 0.95\textwidth]{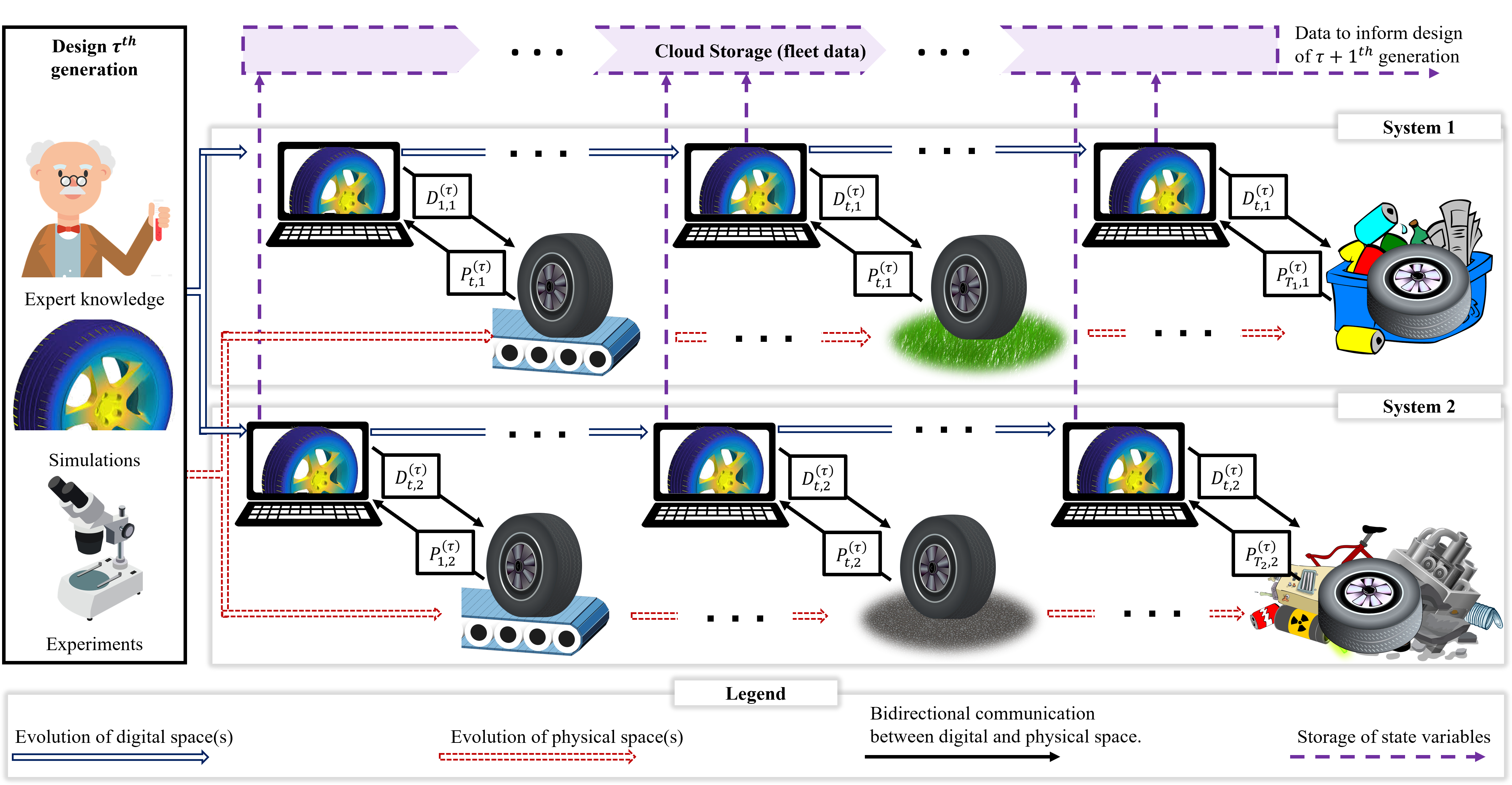}
\end{center}
\caption{\rev{\textbf{Visualization of the digital twin-inspired approach for the design of a vehicle tire.} During the design phase, fleet data (i.e., data from all previous systems) can be used to make nominal design and operation decisions. Conversely, during a system's operation life data can be used to make operation decisions for a specific system as they are exposed to unique operating conditions (e.g., different operator decisions and environmental conditions).}}
\label{TireExampleFig} 
\end{figure}

An example of the application of the digital twin concept for the design of a system is car tires, as visualized in Figure~\ref{TireExampleFig}. \rev{Car tires can benefit from the digital twin concept as controlling tire pressure require bidirectional communication between the physical systems and the digital system to improve fuel economy, extend tire life, improve handling, and reduce risks \cite{Velupillai2007,Kowalewski2004}}. \rev{Throughout the different stages, the state of a tire can be inferred and monitored during its life cycle.} For example, during the operation of a car tire, sensors can be used to measure, tire pressure, tire velocity, tire temperature, total mileage covered by the tire, and tire wear rate (i.e., variables $\textbf{P}_t$). \rev{The collected data then functions as input to the digital system to provide an approximation of the physical state of the tire.} For example, simulation models can use the sensor data as input to predict an additional set of state variables such as strain energy and the evolution of tears inside the tire (i.e., variables $\textbf{D}_t$). Note that the simulation models can be physics-based, data-driven or a combination of the two \cite{Mao2020}. The joint set of the sensors and simulation data provide the approximation of the state of a specific tire for a specific vehicle (i.e., $\textbf{s}_t=\textbf{P}_t \cup \textbf{D}_t$). Note that this implies that the state of a car tire is context-dependent (i.e., driving behavior, weather, and road conditions influence the evolution of a tire's state). \rev{This observation has been visualized by the evolution of System 1 and System 2 in Figure~\ref{TireExampleFig}. For example, during the manufacturing stage, variations in microstructural features can result in a difference in system performance and the effect of these variations can exacerbate over time.} However, the state representation will typically be a high-dimensional data source that is difficult to interpret. Consequently, a reward function $r_t(\cdot)$ is required that uses data from physics-based models, physical experiments, and tire/car-specific sensor data to enable quantitative comparison between the merit of different states at a specific time $t$. The reward function could consider a wide variety of criteria that can include the remaining tire life, driving comfort, tire economy, and environmental impact. Finally, we require a system dynamic $\Gamma_t(\cdot)$ that enables the prediction of future states $\textbf{s}_{t+1}, \textbf{s}_{t+2}, \ldots$ given a set of control policies $\mu(\cdot)$, weight their relative measure, and then select the best policy. Examples of control policies $\mu(\cdot)$ for a car tire include the regulation of tire pressure, vehicle velocity/acceleration, and when to replace the tire. While the current description of the car tire system focuses on operation decisions, it should be observed that its formulation also allows for the consideration of design decisions $\textbf{x}$ (e.g., tire thickness, thread depths, and material), and operation decisions during different stages of the life cycle (e.g., manufacturing process conditions and disposal). This example will be used throughout this paper to contextualize different aspects of the digital twin concept for the design of systems as introduced in this section.

\subsection{Capabilities of a digital twin for System Design}
\label{capabilitiesection}
From the above discussion, we can identify the following set of necessary capabilities when using the digital twin concept for the design of systems. 
\begin{enumerate}
    \item \textbf{Online monitoring and control (i.e., collection of state $\textbf{s}_t$, and control data $\textbf{u}_t$):} A primary function of the digital twin concept is the ability of the digital system to use data collected from the physical system to provide monitoring of the state of the physical system (e.g., in the car tire example the tire pressure is measured continuously and adjusted to improve performance). This is a crucial capability as this provides online insight into the state of a specific system as opposed to the information of a fleet of systems. Moreover, through online monitoring, it becomes possible to make system-specific decisions by providing a control signal to the physical state. 
    \item \textbf{Forward prediction (i.e., the dynamic $\Gamma$):} Forward prediction enables a system-specific estimation of future performance. This capability improves the effectiveness of the online monitoring and control capability, as it enables the consideration of the impact that decisions have at future time steps of a system's life cycle. (e.g., in the car tire example the endurance strength of the tire can be predicted by utilizing temperature and wear rate as measured through the sensors placed on the physical system). Observe that this highlights the importance of uncertainty quantification of future states so that operation decisions can be made that promote either robustness or reliability. 
    \item \textbf{Multi-modal data fusion (i.e., improve the fidelity of $\Gamma(\cdot)$ and $f(\cdot)$):} This capability enables the integration of data collected from the physics-based simulation models with the data collected from the specific system or fleet of systems. The physics-based models provide only an approximation of the true physical system and as such data can be used to improve its fidelity. This capability enables the predictions of the digital system to converge to the physical system and as such approximate the full potential of the digital twin concept (e.g., in the car tire example the bias in the physics-based endurance strength model can be accounted for by comparing data from physical experiments). In this scenario, data from previous systems can also be used to improve the predictive fidelity of the models used in the design phase.
    \item \textbf{Online updating of the digital system (i.e., update $\Gamma(\cdot)$ at times $t$):} This capability enables the fidelity of the forward prediction capability to improve over time and to be system-specific. Consequently, this improves the effectiveness of the operation decisions $\textbf{u}_t$. The updating capability can be achieved by comparing the sensor data with the predicted state of the system that comes from the forward prediction capability $\Gamma(\cdot)$. Through this comparison, the digital system can be updated throughout the life cycle of a specific system (e.g., in the car tire example, the endurance strength model continuously improves in fidelity as more sensor data is acquired). This will improve the predictive properties of the forward prediction capability and improve the control decisions. In other words, online updating of the digital system is specific to a system whereas the multi-modal data fusions capability can also inform decisions of a fleet of systems. 
    \item \textbf{Dimension Reduction of State Representation (i.e., minimize $d_s + p_s$):} In order to make Equation~\ref{POMDP_approx2} tractable, it is required that we have a low dimensional representation of the physical state $s_t$. Consequently, this requires the exclusion of large amounts of information that will inevitably result in a discrepancy between the predicted future states and the observed future states (e.g., in the car tire example we cannot measure the temperature field throughout the tire, but have to rely on a single point measurement). Note that this implies that dimension reduction of the state representation involves the appropriate selection of simulation model output as well as the selection of sensors and their placement. Consequently, this will require an approach that can be used to reduce the dimension of the state representation while minimizing the negative consequences that this has on the fidelity of the state transition function $\Gamma(\cdot)$.  
    \item \textbf{Exploratory data collection (i.e., account for the fidelity of the system dynamic $\Gamma(\cdot)$ in the objective function $f(\cdot)$):} When invoking the digital system to inform decisions, it could happen that some areas of the space of admissible designs and control decisions have not been explored. Consequently, the prediction fidelity in these areas is low and it is impossible to confidently conclude that these decisions are inferior (e.g., in the car tire example the system dynamic $\Gamma(\cdot)$  cannot accurately predict if the driving distance can be improved by reducing the tire pressure to 30 psi if it only has data from physical experiments, simulations, and operation decisions in the range of 50 psi to 100 psi). An additional capability of the digital twin concept for the design of systems would thus be to enable the exploration of decisions that have a large degree of prediction uncertainty. This capability will improve the global fidelity of the digital system to have an improved value proposition over its entire life cycle. Note that this capability necessitates the ability to quantify the predictive uncertainty of the digital system to balance the predicted average merit of a decision with its uncertainty.
    \item \textbf{State Reward Function Approximation (i.e., the formulation of $r(\cdot)$):} In the optimization of Equation~\ref{POMDP_approx2} we require a formulation for the reward function of the state of a system. The complexity of this function depends on the intended purpose of the system. For example, the length of a crack in an aircraft wing is used as the reward function in \cite{Mahadevan2017}. However, in most cases, the formulation of a reward function will not be as straightforward as it involves the trade-off between different objectives, or the need to balance the disparate interests of a group of individuals interacting with the system. This capability enables the designer of a system to autonomously measure the merit of design and operation decisions. 
    \item \textbf{Uncertainty Quantification and Propagation (i.e., predict the uncertainty in $f(\cdot)$):} The optimization problem stated in Equation~\ref{POMDP_approx} requires knowledge about the uncertainty in the objective function. \rev{Note that in conjunction with capability 6, it is necessary to make a distinction between epistemic and aleatoric sources of uncertainty. The reason for this is that the acquisition of new data points can only be expected to reduce epistemic uncertainty (e.g., model uncertainty), as aleatoric uncertainty is intrinsic to the acquired data.} Consequently, an important capability of establishing a digital system is the quantification of uncertainty sources and the propagation to the objective function (e.g., in the car tire example the digital system does not know the environmental conditions under which the tire will be exposed in the future and thus needs to account for this source of uncertainty). \rev{Moreover, it should be noted that through the updating capability newly observed data points can be used to reduce uncertainty, (e.g., physical data is used to validate and quantify the uncertainty of a simulation model).} Uncertainty quantification enables a designer to make a trade-off between the mean performance and variability when deciding to make system design and/or operation decisions. 
\end{enumerate}

\section{Status of the Digital Twin Concept for System Design}
\label{CurrentState}

In this section, we discuss some of the literature relevant to the use of digital twins for the design of systems. We list a set of representative papers and compare them with the capabilities identified in Section~\ref{capabilitiesection}. \rev{It was found that most of the available literature integrated three or four capabilities while specifically focusing on one. Consequently, in Table~\ref{LiteratureTable} the research papers are clustered based on their primary capabilities (vertical axis of the table). In addition, any secondary capabilities of the papers reported in Table~\ref{LiteratureTable} are summarized in columns three to eleven.}

\begin{table}[ht]
\caption{Comparison of literature on the digital twin concept to the capabilities identified in Section~\ref{capabilitiesection}.}
\centering 
\begin{tabular}{p{40mm} p{20mm} p{5mm} p{5mm} p{5mm} p{5mm} p{5mm} p{5mm} p{5mm} p{5mm}}
\toprule
 & & \multicolumn{8}{c}{Secondary capability number}\\
 Primary capability & Reference & 1 & 2 & 3 & 4 & 5 & 6 & 7 & 8\\
\midrule

\multirow{3}{40mm}{1. Online monitoring and control} 
    & \cite{zhou2019digital}& X  & X &   &  X &   & X  &   &   \\
    & \cite{cronrath2019enhancing} &  X  &   &   & X  &   &   & X  & X \\
    & \cite{pan2021bim} & X & X  &   & X  &   &   &  &  \\
    \midrule

\multirow{6}{40mm}{2. Forward prediction} 
    & \cite{jiang2021industrial} & X & X &   & X &   &   &   &    \\
    & \cite{brockhoff2021process} &   & X &   &   &   &   &   &   \\
    & \cite{wunderlich2021digital} &  X  & X  &   & X  &   &   &   &   \\
    & \cite{jaensch2018digital} &    & X  & X  &   &   &   &   &   \\
    & \cite{suhail2021trustworthy} &   & X  &   &   &   &   &   & X \\
    & \cite{tuegel2011reengineering} & X & X & X  & X  &   & X  &  & X \\\midrule

\multirow{3}{40mm}{3. Multi-modal data fusion} 
    & \cite{liu2018role} & X  & X & X  &   &   &   &   &   \\
    & \cite{xiang2018digital} &    & X  &X   &   &   & X  &   & X  \\
    & \cite{erikstad2017merging} & X & X  & X  & X  &   &   &   &  \\\midrule

\multirow{3}{40mm}{4. Online updating of the digital system} 
    & \cite{huang2021digital} &  X  & X  &   &  X &   & X  &   &   \\
    & \cite{koulamas2018cyber} & X  & X  & X  & X  & X  &   &   &  X \\
    & \cite{ritto2021digital} & X & X  & X  & X  &   &   &  & X \\\midrule

\multirow{2}{40mm}{5. Dimension Reduction of State Representation} 
    & \cite{hartmann2018model} &    &   & X  &   & X  &   &   &   \\ 
    & \cite{garg2021multi} &    & X  &   &   & X  &   &   &   \\\midrule

\multirow{3}{40mm}{6. Exploratory data collection} 
    & \cite{gohari2019digital} &    & X  &   &   &   & X  &   &  X \\
    & \cite{barkanyi2021modelling} & X &   & X  & X  &   & X  &  &  \\
    & \cite{macchi2018exploring} & X &   &   & X  &   & X  &  &  \\\midrule

\multirow{2}{40mm}{7. State Reward Function Approximation} 
    & \cite{sun2020adaptive} &    & X  &   &   &   &   & X  & X  \\
    & \cite{eirinakis2020enhancing} &   & X  & X  & X  &   & X  & X  &   \\\midrule

\multirow{3}{40mm}{8. Uncertainty Quantification and Propagation} 
    & \cite{woodcock2020uncertainty} &  X  & X  &   &  X &   &   & X  &  X \\
    & \cite{gohari2019digital} & X  & X  &   & X  &   &   &   &  X \\
    & \cite{aversano2019application} & X &   & X  & X  &   &   & X & X \\

\bottomrule
\end{tabular}
\label{LiteratureTable}
\end{table}


\rev{From the literature on monitoring and control (capability 1), we can identify two types of approaches. In the first approach, sensors are used as input to a model to generate insight into a specific system that is then used by a human to make control decisions \cite{zhou2019digital,pan2021bim}. For example, in \cite{zhou2019digital} a feed-forward neural network is used to establish an online analysis tool to make power grid management decisions, and in \cite{pan2021bim} a framework that integrates building information models, with the internet of things, and data mining is used to help inform project management decisions. Conversely, in the second approach, data is collected through sensors and used in the digital system to make autonomous decisions (i.e., no human is needed to make decisions). For example, in \cite{cronrath2019enhancing} reinforcement learning is proposed as an approach to learning the system dynamic $\Gamma(\cdot)$ and reward function $r(\cdot)$ to make autonomous control decisions. The ability to make autonomous control decisions is an important functionality when using the digital twin concept for the design of systems, but it is not sufficient as it cannot readily be extended to also include design decisions.}

\rev{Part of autonomous decision-making is the construction of the system dynamic $\Gamma(\cdot)$ that predicts how the system will evolve. Through forward prediction of the state $\textbf{s}_t$ (capability 2), it becomes possible to make non-myopic decisions that consider the remaining life-cycle of the system as opposed to greedy decisions that only consider the state of the system at the next time step. The importance of this capability has been argued in \cite{jiang2021industrial,brockhoff2021process}. Examples of specific methods for forward prediction of the system state include Bayesian machine learning \cite{jiang2021industrial}, dynamic neural networks \cite{wunderlich2021digital}, and physics-informed machine learning \cite{tuegel2011reengineering}. Examples of successful implementation of forward state prediction include power electronic converters \cite{wunderlich2021digital}, manufacturing \cite{jaensch2018digital}, and maintenance \cite{suhail2021trustworthy,tuegel2011reengineering}.}

\rev{Forward prediction of the system state is a prerequisite for the ability to quantitatively measure the relative merit of competing design and operating decisions through a state reward function $r(\cdot)$ (capability 7). In \cite{sun2020adaptive}, a framework for federated learning and deep reinforcement learning is introduced to analyze power consumption. Federated learning is a data driven method that enables statistical learning on multiple decentralized systems. Jointly, the system dynamic and reward function enable the digital system to have information on the physical characteristics and behavior of some of the physical elements, an emerging capability that the authors of \cite{eirinakis2020enhancing} refer to as the ``cognitive digital twin''.} 

\rev{Additional capabilities to improve the effectiveness of the digital twin concept for the design of systems is the need to fuse multi-modal data \cite{xiang2018digital} (capability 3). Examples of this are the integration of sensor data with simulation data \cite{liu2018role} and the weighted sum of multiple models \cite{erikstad2017merging}.} \rev{The ability to fuse data from multiple sources of fidelity which can be achieved by multi-fidelity modeling can greatly improve the predictive accuracy of the system dynamic $\Gamma(\cdot)$ and reward function $r(\cdot)$ and thus the effectiveness of the rendered decisions. A challenge with multi-modal data fusion is updating the digital system to mimic a specific physical system \cite{koulamas2018cyber} (capability 4). Note that this is different from the initial digital system that is constructed during the design phase and that is the same for the entire fleet of systems. Specifically, in \cite{koulamas2018cyber} a framework is presented that integrates the internet of things with big data principles, ambient intelligence, and robotics to enable online learning.}

\rev{Another aspect of effective decision-making is uncertainty quantification \cite{aversano2019application} (capability 8), which has been achieved through Monte Carlo-based sampling \cite{woodcock2020uncertainty} or the more efficient surrogate-based approaches like Gaussian processes \cite{aversano2019application}. Updating the digital system and uncertainty quantification with forward prediction can be a computational and time-intensive challenge that limits practical implementation. Examples of models that have been applied successfully for this purpose are random forests \cite{ritto2021digital}, and edge intelligence \cite{huang2021digital}. Edge intelligence is an approach that can accelerate model updating by dividing the process in multiple processes of reduced computational cost and then performing these simulations in a decentralized manner (i.e., reduced computational cost through parallel computing). However, note that not all systems are amenable for edge intelligence as not all computational processes can be performed in parallel.}

\rev{Model updating can be further accelerated through dimension reduction (capability 5), as acknowledged in \cite{hartmann2018model}. An additional advantage is that dimension reduction reduces the need for data storage \cite{garg2021multi}. Specific examples of dimension reduction within the context of a digital twin concept include the Krylov subspace method \cite{hartmann2018model}, and data duplicity elimination for intrinsically noisy data sources  \cite{garg2021multi}.} \rev{In addition, updating the digital system, enables intelligent exploratory data collection that involves a compromise in the effectiveness of immediate operation decisions to provide long-term operational benefits \cite{koulamas2018cyber,ritto2021digital} (capability 6). This capability can be further enhanced through uncertainty quantification to balance the need for exploitation with exploration \cite{huang2021digital}. While the individual capabilities of the digital twin concept have been explored without explicit consideration of system design, we find that they all fit in the general framework as presented in Section~\ref{DTandDesign}}. 

In addition to the above observations, we find that no work presents a framework that integrates all the capabilities identified in Section~\ref{capabilitiesection}. Additionally, it can be observed that there is a lack of research on the quantification and propagation of uncertainty sources. Not accounting for the various sources of uncertainty (e.g., data, model, and prediction uncertainty) limits the quality of the decisions that are rendered based on the information provided by the digital state. In addition, the state-reward function approximation is not used widely in models of digital twins for systems design. \rev{While the state reward function can be relatively straightforward for the design of a product with a single objective, its formulation for the design of a system with multiple objective can be complex in nature, and computationally expensive}. More details on the specifics of these challenges are given in Section~\ref{Future}. In short, there are no research papers that jointly implement all the capabilities, this observation in combination with the potential benefits delineated in Section~\ref{DTandDesign} indicates that there are new and worthwhile research areas to be explored in the use of the digital twin concept for the design of systems.

\section{Discussion on Research Opportunities and Relevant Methods}
\label{Future}

From Table~\ref{LiteratureTable} we observed that the full potential of the digital twin concept for the design of systems, as described in this paper, has not yet been achieved. The primary reason for this is that no method integrates all capabilities. Specifically, we find that the following three research areas warrant further investigation: (i) fusion of multi-modal data sources to enable forward prediction of future states, (ii) modeling of the state reward function to measure the relative merit of competing design and operation decisions, and (iii) the integration of all capabilities to provide tractable design and operation support.

\subsection{Avenues for Further Scientific Inquiry}
In this subsection, we provide a detailed introduction to the problem-specific intricacies of the three research areas and use this to formulate a set of research questions. 

\begin{enumerate}
\item \textbf{Fusion of multi-modal data sources to enable forward prediction of future states:} Autonomous decision-making places stringent demand on the required computational resources as it involves the fusion of multi-modal data to make predictions of future states when optimizing Equation~\ref{POMDP_approx2}. This process quickly becomes intractable when directly using physics-based simulation models, and thus motivates the use of multi-modal data fusion methods \cite{Chen2017}. Critical to the accuracy of these statistical data fusion models is the number of available samples and the dimensionality. Specifically for the design of systems, the representation of the state $\textbf{s}_t$ is not a straightforward task. The complexity of the physical system can be observed through a set of carefully placed sensors; however, this only provides a discrete representation of a continuous system \cite{Willcox2021}. Consequently, when this information is used to approximate the state of a physical system through a set of computational models, a bias will be observed \cite{Kennedy2001}. Consequently, a systematic approach to reducing this bias through the allocation of a sparse but representative set of sensors and simulation models becomes important. With that in mind, we can identify the following set of research questions;
\begin{enumerate}
    \item How to fuse multi-modal data sources to mimic current and predict future states of a physical system to enable decision-making? 
    \item How can we achieve dimension reduction of the state representation while maintaining the quality of design and operation decisions?
    \item How to quantify the various sources of uncertainty and how to propagate their effects on the objective function in Equation~\ref{POMDP_approx2}?
\end{enumerate}
The advantage that can be achieved by addressing the above objectives is that it will improve the accuracy of the system dynamic while requiring fewer experimental samples to do so. Consequently, the system can make decisions that will have an improved expected reward. Moreover, reducing the dimensionality of the system will reduce the computational overhead, and as such design and operation decisions can be made in shorter time intervals.

\item \textbf{Modeling of the state reward function to measure the relative merit of competing design and operation decisions:} The purpose of the state reward function is to compare the relative merit of competing decisions at a specific point in time $t$ and needs to reflect the preferences of the system's stakeholders. While this can be straightforward for a design that has a unitary metric of merit (e.g., failure stress in an aircraft wing \cite{Willcox2021} or crack length in an aircraft wing \cite{Mahadevan2017}), the typical scenario will be that the merit of a system is measured through multiple criteria. We will continue our discussion by going into multi-objective optimization, but the reader should be aware that the state reward function cannot directly be applied as the objective function due to the recursive formulation of Equation~\ref{POMDP_approx2}. Multi-objective optimization has a rich body of literature and can typically be divided into two types of approaches; (i) finding a Pareto frontier of optimal solutions and then choosing among those \cite{Iyer2019}, and (ii) optimizing a weighted sum of objective functions \cite{Hazelrigg1998}. However, most of these approaches are only appropriate for problems involving a relatively small number of criteria ($\leq 3$). In contrast, the merit of a system can often be measured over a comparatively much larger number of criteria. For the car tire example, a designer could be interested in fuel economy, maintenance cost, environmental impact, driving comfort, and average mileage. It is difficult to select a decision on a Pareto frontier or to weigh the relative importance of dimensions of merit in a space with four or more dimensions. More complicated still is the consideration that many systems are designed for a plurality of stakeholders (e.g., car users, manufacturers, mechanics, designers, and dealers) that hold divergent preferences, a problem that is known as preference modeling \cite{Cui2022}. Note that the relative importance of the dimensions of merit is weighed differently by each individual and that this cannot be mapped into a single preference without violating conditions of logical decision-making \cite{Yu2012,Arrow1950,Beek2022}. While preference modeling addresses this problem by using consumer purchasing data to identify the design features and their associated levels that are predicted to maximize sales, it does not account for other stakeholders nor does it account for the evolution of a system over its life cycle. For example, a mechanic replacing a tire might have a different preference when it comes to the design of the tire as opposed to a consumer. In addition, the mechanic's preference will not be reflected in their purchasing behavior/data. Consequently, new methods are required to make decisions about the design and operation of a system that consider the interests of all the stakeholders interacting with the system. From this observation, the following two research questions can be identified;
\begin{enumerate}
    \setcounter{enumii}{3}
    \item How to formulate the state reward function, considering a large number of criteria, to account for the effectiveness of design and operation decisions over the life cycle of a system?
    \item How can the divergent interests of stakeholders be considered when making design and operation decisions?
\end{enumerate}
Through scientific inquiry into the above two research questions, a deeper understanding of high-dimensional and group decision-making will be obtained. Consequently, a more accurate and logically consistent measure of merit for competing decisions will be obtained to improve the quality of the rendered decisions. In addition, systematically considering the interests of the stakeholders will improve the effectiveness of the rendered decisions as they will enjoy broader social support. Finally, establishing more accurate state reward functions will improve the degree to which decisions can be made autonomously as part of the digital system.

\item \textbf{Integration of all capabilities to provide tractable design and operation support:} Finally, the integration of all capabilities of the digital twin concept for the design of systems requires additional scientific inquiry. The full potential, as described in Section~\ref{DTandDesign}, has not yet been achieved as no framework has successfully combined all the identified capabilities. The objective of integrating all capabilities itself has a set of challenges. Specifically, we forecast that the computational overhead will become a limiting factor. For example, online updating of the digital systems, uncertainty quantification, uncertainty propagation, and optimization of the objective function are in themselves computationally intensive tasks. Considering these tasks jointly, in one framework, is likely to result in prohibitive computational cost. Even the reduction of the dimensionality of the state will not sufficiently reduce the computational cost of Equation~\ref{POMDP_approx} to enable decision-making during a system's life cycle. Consequently, additional effort is required to reduce the computational cost of integrating all capabilities. The authors believe that this can be achieved through improved computational algorithms and simplifying assumptions. Consequently, this gives rise to the following set of research questions;
\begin{enumerate}
    \setcounter{enumii}{5}
    \item How can we combine the capabilities, identified in this paper, for design and operation decision-making? 
    \item What level of digital system fidelity is tolerable when making predictions and decisions?
\end{enumerate}
The advantages of combining all capabilities of the digital twin concept for the design of systems will enable the realization of systems with improved value. Specifically, this will enable the efficient use of data to improve model fidelity and the reliability of the associated design and operation decisions. More important still, is that the digital twin concept for the design of systems will enable the joint consideration of design operation decisions.
\end{enumerate}

\subsection{Relevant Statistical Tools}
While the solution strategies for the above-identified research directions are still to be defined, it is clear that they will require the use of statistical tools to enable data-informed decision-making. Consequently, in this subsection, we will go over the identified research directions and introduce a set of statistical tools that can prove fundamental to answering their associated research questions.

\begin{enumerate}
    \item \textbf{Relevant tools for the fusion of multi-modal data sources to enable forward prediction of future states:} \indent The dynamic $\Gamma(\cdot)$ of the digital system will in many scenarios be required to predict the state of the system at future points in time \cite{zhang2010moodcast}. Note that the system dynamic $\Gamma(\cdot)$ is different from the reward function $r(\cdot)$ in that it needs to provide a forward prediction of the system state as opposed to providing a relative measure of merit. First, we discuss the scenario where a data-driven model is required to predict temporal and continuous response variables (e.g., in the car tire example we have the tire temperature and residual casing stress as real-valued and temporal variables). Some relevant models that can be used in this context include autoregressive models \cite{Hannan1986}, dynamic regression \cite{shaw1997dynamic}, neural network-based models \cite{hwang2009dynamic,zaremba2014recurrent}, and Gaussian process-based models \cite{Conti2009, zhao2011metamodeling}. A specific advantage offered by Gaussian process-based models is that they provide uncertainty quantification that can be leveraged for efficient optimization (i.e., Bayesian optimization) \cite{zhang2017digital,dehghanimohammadabadi2021simulation}. 
    
    In addition to continuous state variables, the dynamic $\Gamma(\cdot)$ can be required to predict a discrete future state. For example, in the car tire problem the prediction of whether the tire will fail at times $t\in \left\{1,\ldots,T\right\}$ (i.e., yes or no?). In addition to the systems dynamic, the policy $\mu(\cdot)$ can be required to predict a discrete set of operating decisions given the state of a system $\textbf{s}_t$ (e.g., should the tire be retreaded?). It thus follows that when we require a model to predict the discrete state of a system $\textbf{s}_t$ or map the state of a system into a discrete decision (i.e., $\mu(\textbf{s}_t)$), then we are dealing with a classification problem. Addressing this type of problem requires recognition, understanding, and grouping of discrete objects or data at times $t\in \left\{1,\ldots,T\right\}$\cite{rafiei2017new, barthelmey2019dynamic}. For this purpose, a promising class of classification models are random forests as they have strong performance in classification, can be compatible with temporal data, and provide uncertainty quantification of the predicted class \cite{xu2017implementation}.
    
    \item \textbf{Relevant tools for modeling the state reward function to measure the relative merit of competing design and operation decisions:} While it is an open research question of how the state reward function $r(\cdot)$ should be constructed, its formulation and validation will require data (experimental and simulation). In the case of the digital twin concept for the design of systems, multi-modal data is collected with the aid of sensors in real-time \cite{aydemir2020digital}. Typical multi-modal fusion methods, (e.g., calibration and bias correction \cite{Kennedy2001} cokriging \cite{Khatamsaz2021}, nonhierachrical fusion \cite{Chen2016}, and latent map data fusion \cite{Bostanabad2022}), enable the integration and interpolation of multiple sources of noisy data \cite{gong2022data}. While these methods work well for low dimensional data, the design of systems typically requires statistical models that can handle high dimensional data \cite{erikstad2017merging}. Specifically, the high dimensionality of the data can come from several factors; (i) the data can include images and property fields (e.g., the stress fields inside the tire as predicted by a simulation model), and (ii) some systems include a large number of sensors (e.g., an aircraft or truck). Consequently, this requires a statistical method that is compatible with high-dimensional inputs and outputs. With that in mind, Bayesian networks provide a promising approach and have the additional advantage of being able to update the model parameters as new data becomes available (i.e., improve model fidelity with incoming data)\cite{Chen2012}.  
    
    The data received from sensors, simulation models, and physical experiments typically manifest various types of uncertainty \cite{chakraborty2021role, Kennedy2001}. Consequently, the reward function should account for these sources of uncertainty to facilitate decision-making. While the above models can be used to integrate and interpolate between the multi-modal data, they do not provide a systematic treatment of the sources of uncertainty and their effect on the predicted responses. The complexity of this task should not be underestimated as establishing a digital system will require the integration of a great many statistical tools that each have their specific source of uncertainty. For example, a typical approach when incomplete data is observed (e.g., when sensors fail to give information about the pressure of the tire as a consequence of an internet outage) is to use synthetic data to replace it \cite{zotov2020towards}. However, while synthetic data can be used to keep the digital system operational, it also introduces a new source of uncertainty that needs to be quantified and propagated \cite{xu2022design, tuegel2012airframe}. Some statistical tools that have historically proven useful for these types of tasks include Gaussian process models \cite{Beek2021}, Bayesian neural networks \cite{Kononenko1989}, Bayesian networks \cite{Mahadevan2017} polynomial chaos expansion \cite{Sepahvand2010}, and quantile regression \cite{Koenker2001}. In the car tire example, the velocity sensor has an error rate of $\pm 1$ percent that needs to be accounted for when making design and operation decisions.

    \item \textbf{Relevant tools for integration of all capabilities to provide tractable design and operation support:} Potentially the biggest challenge that needs to be addressed to realize the advantages that the digital twin concept holds for the design of systems is tractability. The individual capabilities of the digital system can be computationally intensive, making their joint integration more complicated still. Computational advantages can be achieved by decoupling the constructs of the digital system so that they can be performed through parallel computing \cite{Zhou2019}. Such an approach typically lends itself to one of three approaches; bagging, boosting, and stacking \cite{syarif2012application}. In bagging, multiple models are trained in parallel on a subset of the training data and then combined through averaging. In boosting, models learn sequentially in and adaptively place more weight on poorly predicted samples to reduce prediction bias. In stacking, different models are trained in parallel and are then combined by spatially changing the weights on each model. In contrast, algorithmic advancements will require research into the specific methods, so that their unique properties can be leveraged to improve the performance of the digital system. 
\end{enumerate}

\section{Concluding Remarks}
\label{Conlusions}
In this paper, we have presented a digital twin-inspired perspective on the design of systems. Specifically, we introduced a system-agnostic mathematical formulation for how the digital twin concept can be used to improve the value proposition of systems by enabling the joint optimization of design and operation decisions. This formulation involves two primary aspects; (i) design decisions that define the materials, dimension, and configuration of a system and that are fixed throughout a system's life cycle, and (ii) autonomous operation decisions that get updated based on a system's specific state and usage conditions. We continued our discussion by comparing available literature on the digital twin concept for the design of systems with the identified capabilities. From this analysis, we observed that no method exists that integrates all capabilities stated in this paper, and thus concluded that additional scientific inquiry into the use of the digital twin concept for the design of systems is warranted. Subsequently, we highlighted a set of research areas for future scientific inquiry that will provide a valuable contribution to the design of systems. These directions include; (i) dimension reduction of state representation (i.e., the parameters used to characterize the digital, and physical state), (ii) the formulation of a state reward function, \rev{ (iii) online updating of the digital system through forward prediction}, and (iv) the integration of the required capabilities to realize the potential that the digital twin concept holds for the design of systems, as described in this paper. \rev{Under each of the general research directions, we have identified a set of research questions and introduced a set of statistical tools that can prove useful in answering these questions.} From this exercise we formulated the following key takeaways;
\begin{enumerate}
    \item design and operation decisions require quantitative comparison between their merit on different stages in a system's life cycle,
    \item establishing an efficient digital system involves a trade-off between dimension reduction of state representation and the quality of the design and operating decisions, and
    \item integration of the identified capabilities requires significant computational advances to be tractable.
\end{enumerate} 

We argue that, while the digital twin concept has received progressively more interest from the scientific community, its full potential for the design of systems has not yet been achieved. By working towards the aforementioned digital twin-inspired research areas we believe that more advanced design systems can be realized. The advantages that these systems will have are; (i) improved performance through the joint consideration of design and operation decisions, and (ii) improved data efficiency as it will enable the use of data across multiple design generations. By laying out a set of research directions we hope to guide the scientific community toward achieving the lofty goal of using the digital twin concept to enable the designs of systems to address the societal challenges of the future (e.g., the design of advanced data-driven health systems, urban infrastructure systems, and clean water systems to address some of the grand challenges as identified by the national academy of engineering \cite{Nae2019}).

\section*{Acknowledgements}
Grant support from the National Science Foundation future manufacturing research program NSF 2037026 is greatly acknowledged

\section*{Conflict of Interest}
On behalf of all authors, the corresponding author states that there is no conflict of interest.

\section*{Replication of Results}
No results are presented.

\bibliographystyle{ieeetr}
\bibliography{Bibliography}

\begin{thebibliography}{10}

\bibitem{fuller2019digital}
A.~Fuller, Z.~Fan, and C.~Day, ``Digital twin: Enabling technology, challenges
  and open research,'' {\em arXiv preprint arXiv:1911.01276}, 2019.

\bibitem{tao2019digital}
F.~Tao, Q.~Qi, L.~Wang, and A.~Nee, ``Digital twins and cyber--physical systems
  toward smart manufacturing and industry 4.0: Correlation and comparison,''
  {\em Engineering}, vol.~5, no.~4, pp.~653--661, 2019.

\bibitem{son2022past}
Y.~H. Son, G.-Y. Kim, H.~C. Kim, C.~Jun, and S.~D. Noh, ``Past, present, and
  future research of digital twin for smart manufacturing,'' {\em Journal of
  Computational Design and Engineering}, vol.~9, no.~1, pp.~1--23, 2022.

\bibitem{van2021archetypes}
H.~van~der Valk, H.~Ha{\ss}e, F.~M{\"o}ller, and B.~Otto, ``Archetypes of
  digital twins,'' {\em Business \& Information Systems Engineering},
  pp.~1--17, 2021.

\bibitem{haag2019automated}
S.~Haag and R.~Anderl, ``Automated generation of as-manufactured geometric
  representations for digital twins using step,'' {\em Procedia CIRP}, vol.~84,
  pp.~1082--1087, 2019.

\bibitem{Abdulmotaleb2017}
K.~M. Alam and A.~El~Saddik, ``C2ps: A digital twin architecture reference
  model for the cloud-based cyber-physical systems,'' {\em IEEE Access},
  vol.~5, pp.~2050--2062, 2017.

\bibitem{wright2020tell}
L.~Wright and S.~Davidson, ``How to tell the difference between a model and a
  digital twin,'' {\em Advanced Modeling and Simulation in Engineering
  Sciences}, vol.~7, no.~1, pp.~1--13, 2020.

\bibitem{barkanyi2021modelling}
{\'A}.~B{\'a}rk{\'a}nyi, T.~Chov{\'a}n, S.~N{\'e}meth, and J.~Abonyi,
  ``Modelling for digital twins—potential role of surrogate models,'' {\em
  Processes}, vol.~9, no.~3, p.~476, 2021.

\bibitem{schweigert2020conception}
S.~Schweigert-Recksiek, J.~Trauer, C.~Engel, K.~Spreitzer, and M.~Zimmermann,
  ``Conception of a digital twin in mechanical engineering--a case study in
  technical product development,'' in {\em Proceedings of the Design Society:
  DESIGN Conference}, vol.~1, pp.~383--392, Cambridge University Press, 2020.

\bibitem{AIAA2020}
A.~I. of~Aeronautics and Astronautics, ``Digital twin: Defenition \& value.''
  \url{https://www.aiaa.org/docs/default-source/uploadedfiles/issues-and-advocacy/policy-papers/digital-twin-institute-position-paper-(december-2020).pdf},
  12 2020.

\bibitem{rosen2015importance}
R.~Rosen, G.~Von~Wichert, G.~Lo, and K.~D. Bettenhausen, ``About the importance
  of autonomy and digital twins for the future of manufacturing,'' {\em
  Ifac-papersonline}, vol.~48, no.~3, pp.~567--572, 2015.

\bibitem{sharma2020digital}
A.~Sharma, E.~Kosasih, J.~Zhang, A.~Brintrup, and A.~Calinescu, ``Digital
  twins: State of the art theory and practice, challenges, and open research
  questions,'' {\em arXiv preprint arXiv:2011.02833}, 2020.

\bibitem{soderberg2017toward}
R.~S{\"o}derberg, K.~W{\"a}rmefjord, J.~S. Carlson, and L.~Lindkvist, ``Toward
  a digital twin for real-time geometry assurance in individualized
  production,'' {\em CIRP annals}, vol.~66, no.~1, pp.~137--140, 2017.

\bibitem{reuther2019survey}
A.~Reuther, P.~Michaleas, M.~Jones, V.~Gadepally, S.~Samsi, and J.~Kepner,
  ``Survey and benchmarking of machine learning accelerators,'' in {\em 2019
  IEEE high performance extreme computing conference (HPEC)}, pp.~1--9, IEEE,
  2019.

\bibitem{bergquist2001vector}
N.~Bergquist, ``Vector-borne parasitic diseases: new trends in data collection
  and risk assessment,'' {\em Acta tropica}, vol.~79, no.~1, pp.~13--20, 2001.

\bibitem{biesinger2019facets}
F.~Biesinger and M.~Weyrich, ``The facets of digital twins in production and
  the automotive industry,'' in {\em 2019 23rd international conference on
  mechatronics technology (ICMT)}, pp.~1--6, IEEE, 2019.

\bibitem{halemane1983optimal}
K.~P. Halemane and I.~E. Grossmann, ``Optimal process design under
  uncertainty,'' {\em AIChE Journal}, vol.~29, no.~3, pp.~425--433, 1983.

\bibitem{famelis2019managing}
M.~Famelis and M.~Chechik, ``Managing design-time uncertainty,'' {\em Software
  \& Systems Modeling}, vol.~18, no.~2, pp.~1249--1284, 2019.

\bibitem{Magargle2017}
R.~Magargle, L.~Johnson, P.~Mandloi, P.~Davoudabadi, O.~Kesarkar,
  S.~Krishnaswamy, J.~Batteh, and A.~Pitchaikani, ``A simulation-based digital
  twin for model-driven health monitoring and predictive maintenance of an
  automotive braking system,'' in {\em Proceedings of the 12th International
  Modelica Conference, Prague, Czech Republic, May 15-17, 2017}, no.~132 in 1,
  pp.~35--46, Link{\"o}ping University Electronic Press, 2017.

\bibitem{barthelmey2019dynamic}
A.~Barthelmey, E.~Lee, R.~Hana, and J.~Deuse, ``Dynamic digital twin for
  predictive maintenance in flexible production systems,'' in {\em IECON
  2019-45th Annual Conference of the IEEE Industrial Electronics Society},
  vol.~1, pp.~4209--4214, IEEE, 2019.

\bibitem{tuegel2011reengineering}
E.~J. Tuegel, A.~R. Ingraffea, T.~G. Eason, and S.~M. Spottswood,
  ``Reengineering aircraft structural life prediction using a digital twin,''
  {\em International Journal of Aerospace Engineering}, vol.~2011, 2011.

\bibitem{Seshadri2017}
B.~R. Seshadri and T.~Krishnamurthy, ``Structural health management of damaged
  aircraft structures using digital twin concept,'' in {\em 25th aiaa/ahs
  adaptive structures conference}, p.~1675, 2017.

\bibitem{Wei2022}
T.~Wei, A.~van Beek, J.~Hao, H.~Zhang, and W.~Chen, ``Bayesian calibration of
  performance degradation in a gas turbine-driven compressor unit for prognosis
  health management,'' {\em Journal of Engineering for Gas Turbines and Power},
  2022.

\bibitem{Wang2019}
X.~Wang and L.~Wang, ``Digital twin-based weee recycling, recovery and
  remanufacturing in the background of industry 4.0,'' {\em International
  Journal of Production Research}, vol.~57, no.~12, pp.~3892--3902, 2019.

\bibitem{Ayani2018}
M.~Ayani, M.~Ganebäck, and A.~H. Ng, ``Digital twin: Applying emulation for
  machine reconditioning,'' {\em Procedia CIRP}, vol.~72, pp.~243--248, 2018.
\newblock 51st CIRP Conference on Manufacturing Systems.

\bibitem{leng2021digital}
J.~Leng, D.~Wang, W.~Shen, X.~Li, Q.~Liu, and X.~Chen, ``Digital twins-based
  smart manufacturing system design in industry 4.0: A review,'' {\em Journal
  of manufacturing systems}, vol.~60, pp.~119--137, 2021.

\bibitem{ghanem2020probabilistic}
R.~Ghanem, C.~Soize, L.~Mehrez, and V.~Aitharaju, ``Probabilistic learning and
  updating of a digital twin for composite material systems,'' {\em
  International Journal for Numerical Methods in Engineering}, 2020.

\bibitem{guo2020framework}
D.~Guo, S.~Ling, H.~Li, D.~Ao, T.~Zhang, Y.~Rong, and G.~Q. Huang, ``A
  framework for personalized production based on digital twin, blockchain and
  additive manufacturing in the context of industry 4.0,'' in {\em 2020 IEEE
  16th International Conference on Automation Science and Engineering (CASE)},
  pp.~1181--1186, IEEE, 2020.

\bibitem{moser2020mechanistic}
A.~Moser, C.~Appl, S.~Br{\"u}ning, and V.~C. Hass, ``Mechanistic mathematical
  models as a basis for digital twins,'' {\em Digital Twins}, pp.~133--180,
  2020.

\bibitem{barykin2021place}
S.~Y. Barykin, A.~A. Bochkarev, E.~Dobronravin, and S.~M. Sergeev, ``The place
  and role of digital twin in supply chain management,'' {\em Academy of
  Strategic Management Journal}, vol.~20, pp.~1--19, 2021.

\bibitem{karkaria2021computational}
V.~Karkaria, A.~K. Das, A.~Yadav, A.~Sharma, J.~K. Allen, and F.~Mistree, ``A
  computational framework for social entrepreneurs to determine policies for
  sustainable development,'' in {\em International Design Engineering Technical
  Conferences and Computers and Information in Engineering Conference},
  vol.~85390, p.~V03BT03A019, American Society of Mechanical Engineers, 2021.

\bibitem{jaensch2018digital}
F.~Jaensch, A.~Csiszar, C.~Scheifele, and A.~Verl, ``Digital twins of
  manufacturing systems as a base for machine learning,'' in {\em 2018 25th
  International Conference on Mechatronics and Machine Vision in Practice
  (M2VIP)}, pp.~1--6, IEEE, 2018.

\bibitem{zhou2019digital}
M.~Zhou, J.~Yan, and D.~Feng, ``Digital twin framework and its application to
  power grid online analysis,'' {\em CSEE Journal of Power and Energy Systems},
  vol.~5, no.~3, pp.~391--398, 2019.

\bibitem{Cui2020}
T.~Cui, J.~T. Allison, and P.~Wang, ``A comparative study of formulations and
  algorithms for reliability-based co-design problems,'' {\em Journal of
  Mechanical Design}, vol.~142, no.~3, 2020.

\bibitem{kissmer2018}
T.~Kissmer, T.~Potthoff, and S.~Stieglitz, ``Enterprise digital nudging:
  Between adoption gain and unintended rejection,'' in {\em American Conference
  on Information Systems}, Association For Information System (AIS), 2018.

\bibitem{Garcia2019}
M.~Garcia-Sanz, ``Control co-design: an engineering game changer,'' {\em
  Advanced Control for Applications: Engineering and Industrial Systems},
  vol.~1, no.~1, p.~e18, 2019.

\bibitem{Tsui1999}
K.-L. Tsui, ``Robust design optimization for multiple characteristic
  problems,'' {\em International Journal of Production Research}, vol.~37,
  no.~2, pp.~433--445, 1999.

\bibitem{Youn2004}
B.~D. Youn and K.~K. Choi, ``A new response surface methodology for
  reliability-based design optimization,'' {\em Computers \& structures},
  vol.~82, no.~2-3, pp.~241--256, 2004.

\bibitem{Hazelrigg1998}
G.~A. Hazelrigg, ``{A Framework for Decision-Based Engineering Design},'' {\em
  Journal of Mechanical Design}, vol.~120, pp.~653--658, 12 1998.

\bibitem{Beek2021}
A.~van Beek, U.~F. Ghumman, J.~Munshi, S.~Tao, T.~Chien, G.~Balasubramanian,
  M.~Plumlee, D.~Apley, and W.~Chen, ``Scalable adaptive batch sampling in
  simulation-based design with heteroscedastic noise,'' {\em Journal of
  Mechanical Design}, vol.~143, no.~3, 2021.

\bibitem{Velupillai2007}
S.~Velupillai and L.~Guvenc, ``Tire pressure monitoring [applications of
  control],'' {\em IEEE Control systems magazine}, vol.~27, no.~6, pp.~22--25,
  2007.

\bibitem{Kowalewski2004}
M.~Kowalewski, ``Monitoring and managing tire pressure,'' {\em IEEE
  Potentials}, vol.~23, no.~3, pp.~8--10, 2004.

\bibitem{Mao2020}
Z.~Mao, A.~D. Jagtap, and G.~E. Karniadakis, ``Physics-informed neural networks
  for high-speed flows,'' {\em Computer Methods in Applied Mechanics and
  Engineering}, vol.~360, p.~112789, 2020.

\bibitem{Mahadevan2017}
C.~Li, S.~Mahadevan, Y.~Ling, S.~Choze, and L.~Wang, ``Dynamic bayesian network
  for aircraft wing health monitoring digital twin,'' {\em Aiaa Journal},
  vol.~55, no.~3, pp.~930--941, 2017.

\bibitem{cronrath2019enhancing}
C.~Cronrath, A.~R. Aderiani, and B.~Lennartson, ``Enhancing digital twins
  through reinforcement learning,'' in {\em 2019 IEEE 15th International
  Conference on Automation Science and Engineering (CASE)}, pp.~293--298, IEEE,
  2019.

\bibitem{pan2021bim}
Y.~Pan and L.~Zhang, ``A bim-data mining integrated digital twin framework for
  advanced project management,'' {\em Automation in Construction}, vol.~124,
  p.~103564, 2021.

\bibitem{jiang2021industrial}
Y.~Jiang, S.~Yin, K.~Li, H.~Luo, and O.~Kaynak, ``Industrial applications of
  digital twins,'' {\em Philosophical Transactions of the Royal Society A},
  vol.~379, no.~2207, p.~20200360, 2021.

\bibitem{brockhoff2021process}
T.~Brockhoff, M.~Heithoff, I.~Koren, J.~Michael, J.~Pfeiffer, B.~Rumpe, M.~S.
  Uysal, W.~M. Van Der~Aalst, and A.~Wortmann, ``Process prediction with
  digital twins,'' in {\em 2021 ACM/IEEE International Conference on Model
  Driven Engineering Languages and Systems Companion (MODELS-C)}, pp.~182--187,
  IEEE, 2021.

\bibitem{wunderlich2021digital}
A.~Wunderlich and E.~Santi, ``Digital twin models of power electronic
  converters using dynamic neural networks,'' in {\em 2021 IEEE Applied Power
  Electronics Conference and Exposition (APEC)}, pp.~2369--2376, IEEE, 2021.

\bibitem{suhail2021trustworthy}
S.~Suhail, R.~Hussain, R.~Jurdak, and C.~S. Hong, ``Trustworthy digital twins
  in the industrial internet of things with blockchain,'' {\em IEEE Internet
  Computing}, 2021.

\bibitem{liu2018role}
Z.~Liu, N.~Meyendorf, and N.~Mrad, ``The role of data fusion in predictive
  maintenance using digital twin,'' in {\em AIP conference proceedings},
  vol.~1949, p.~020023, AIP Publishing LLC, 2018.

\bibitem{xiang2018digital}
F.~Xiang, Z.~Zhi, and G.~Jiang, ``Digital twins technolgy and its data fusion
  in iron and steel product life cycle,'' in {\em 2018 IEEE 15th international
  conference on networking, sensing and control (ICNSC)}, pp.~1--5, IEEE, 2018.

\bibitem{erikstad2017merging}
S.~O. Erikstad, ``Merging physics, big data analytics and simulation for the
  next-generation digital twins,'' {\em High-Performance Marine Vehicles},
  pp.~141--151, 2017.

\bibitem{huang2021digital}
H.~Huang, L.~Yang, Y.~Wang, X.~Xu, and Y.~Lu, ``Digital twin-driven online
  anomaly detection for an automation system based on edge intelligence,'' {\em
  Journal of Manufacturing Systems}, vol.~59, pp.~138--150, 2021.

\bibitem{koulamas2018cyber}
C.~Koulamas and A.~Kalogeras, ``Cyber-physical systems and digital twins in the
  industrial internet of things [cyber-physical systems],'' {\em Computer},
  vol.~51, no.~11, pp.~95--98, 2018.

\bibitem{ritto2021digital}
T.~Ritto and F.~Rochinha, ``Digital twin, physics-based model, and machine
  learning applied to damage detection in structures,'' {\em Mechanical Systems
  and Signal Processing}, vol.~155, p.~107614, 2021.

\bibitem{hartmann2018model}
D.~Hartmann, M.~Herz, and U.~Wever, ``Model order reduction a key technology
  for digital twins,'' in {\em Reduced-order modeling (ROM) for simulation and
  optimization}, pp.~167--179, Springer, 2018.

\bibitem{garg2021multi}
A.~Garg and B.~K. Panigrahi, ``Multi-dimensional digital twin of energy storage
  system for electric vehicles: A brief review,'' {\em Energy Storage}, vol.~3,
  no.~6, p.~e242, 2021.

\bibitem{gohari2019digital}
H.~Gohari, C.~Berry, and A.~Barari, ``A digital twin for integrated inspection
  system in digital manufacturing,'' {\em IFAC-PapersOnLine}, vol.~52, no.~10,
  pp.~182--187, 2019.

\bibitem{macchi2018exploring}
M.~Macchi, I.~Roda, E.~Negri, and L.~Fumagalli, ``Exploring the role of digital
  twin for asset lifecycle management,'' {\em IFAC-PapersOnLine}, vol.~51,
  no.~11, pp.~790--795, 2018.

\bibitem{sun2020adaptive}
W.~Sun, S.~Lei, L.~Wang, Z.~Liu, and Y.~Zhang, ``Adaptive federated learning
  and digital twin for industrial internet of things,'' {\em IEEE Transactions
  on Industrial Informatics}, vol.~17, no.~8, pp.~5605--5614, 2020.

\bibitem{eirinakis2020enhancing}
P.~Eirinakis, K.~Kalaboukas, S.~Lounis, I.~Mourtos, J.~M. Ro{\v{z}}anec,
  N.~Stojanovic, and G.~Zois, ``Enhancing cognition for digital twins,'' in
  {\em 2020 IEEE International Conference on Engineering, Technology and
  Innovation (ICE/ITMC)}, pp.~1--7, IEEE, 2020.

\bibitem{woodcock2020uncertainty}
J.~Woodcock, C.~Gomes, H.~D. Macedo, and P.~G. Larsen, ``Uncertainty
  quantification and runtime monitoring using environment-aware digital
  twins,'' in {\em International Symposium on Leveraging Applications of Formal
  Methods}, pp.~72--87, Springer, 2020.

\bibitem{aversano2019application}
G.~Aversano, A.~Bellemans, Z.~Li, A.~Coussement, O.~Gicquel, and A.~Parente,
  ``Application of reduced-order models based on pca \& kriging for the
  development of digital twins of reacting flow applications,'' {\em Computers
  \& chemical engineering}, vol.~121, pp.~422--441, 2019.

\bibitem{Chen2017}
S.~Chen, Z.~Jiang, S.~Yang, and W.~Chen, ``Multimodel fusion based sequential
  optimization,'' {\em AIAA journal}, vol.~55, no.~1, pp.~241--254, 2017.

\bibitem{Willcox2021}
M.~G. Kapteyn, J.~V. Pretorius, and K.~E. Willcox, ``A probabilistic graphical
  model foundation for enabling predictive digital twins at scale,'' {\em
  Nature Computational Science}, vol.~1, no.~5, pp.~337--347, 2021.

\bibitem{Kennedy2001}
M.~C. Kennedy and A.~O'Hagan, ``Bayesian calibration of computer models,'' {\em
  Journal of the Royal Statistical Society: Series B (Statistical
  Methodology)}, vol.~63, no.~3, pp.~425--464, 2001.

\bibitem{Iyer2019}
A.~Iyer, Y.~Zhang, A.~Prasad, S.~Tao, Y.~Wang, L.~Schadler, L.~C. Brinson, and
  W.~Chen, ``Data-centric mixed-variable bayesian optimization for materials
  design,'' in {\em International Design Engineering Technical Conferences and
  Computers and Information in Engineering Conference}, vol.~59186,
  p.~V02AT03A066, American Society of Mechanical Engineers, 2019.

\bibitem{Cui2022}
Y.~Cui, F.~Ahmed, Z.~Sha, L.~Wang, Y.~Fu, N.~Contractor, W.~Chen, and
  S.~Suweis, ``A weighted statistical network modeling approach to product
  competition analysis,'' {\em Complexity.}, vol.~1, 2022.

\bibitem{Yu2012}
N.~N. Yu, ``A one-shot proof of arrow’s impossibility theorem,'' {\em
  Economic Theory}, vol.~50, p.~523–525, 2012.

\bibitem{Arrow1950}
K.~J. Arrow, ``A difficulty in the concept of social welfare,'' {\em Journal of
  Political Economy}, vol.~58, no.~4, pp.~328--346, 1950.

\bibitem{Beek2022}
A.~van Beek, ``A decision-centric perspective on evolving cyber-physical-social
  systems: Effectiveness, group value, and opportunities,'' in {\em 2022
  International Design Engineering Technical Conferences and Computers and
  Information in Engineering Conference}, vol.~1 of {\em 1}, pp.~1--10, 2022.

\bibitem{zhang2010moodcast}
Y.~Zhang, J.~Tang, J.~Sun, Y.~Chen, and J.~Rao, ``Moodcast: Emotion prediction
  via dynamic continuous factor graph model,'' in {\em 2010 IEEE International
  Conference on Data Mining}, pp.~1193--1198, IEEE, 2010.

\bibitem{Hannan1986}
E.~J. Hannan and L.~Kavalieris, ``Regression, autoregression models,'' {\em
  Journal of Time Series Analysis}, vol.~7, no.~1, pp.~27--49, 1986.

\bibitem{shaw1997dynamic}
A.~M. Shaw, F.~J. Doyle~III, and J.~S. Schwaber, ``A dynamic neural network
  approach to nonlinear process modeling,'' {\em Computers \& chemical
  engineering}, vol.~21, no.~4, pp.~371--385, 1997.

\bibitem{hwang2009dynamic}
S.~Hwang, ``Dynamic regression models for prediction of construction costs,''
  {\em Journal of Construction Engineering and Management}, vol.~135, no.~5,
  pp.~360--367, 2009.

\bibitem{zaremba2014recurrent}
W.~Zaremba, I.~Sutskever, and O.~Vinyals, ``Recurrent neural network
  regularization,'' {\em arXiv preprint arXiv:1409.2329}, 2014.

\bibitem{Conti2009}
S.~Conti, J.~P. Gosling, J.~E. Oakley, and A.~O'Hagan, ``Gaussian process
  emulation of dynamic computer codes,'' {\em Biometrika}, vol.~96, no.~3,
  pp.~663--676, 2009.

\bibitem{zhao2011metamodeling}
L.~Zhao, K.~Choi, and I.~Lee, ``Metamodeling method using dynamic kriging for
  design optimization,'' {\em AIAA journal}, vol.~49, no.~9, pp.~2034--2046,
  2011.

\bibitem{zhang2017digital}
H.~Zhang, Q.~Liu, X.~Chen, D.~Zhang, and J.~Leng, ``A digital twin-based
  approach for designing and multi-objective optimization of hollow glass
  production line,'' {\em Ieee Access}, vol.~5, pp.~26901--26911, 2017.

\bibitem{dehghanimohammadabadi2021simulation}
M.~Dehghanimohammadabadi, S.~Belsare, and R.~Thiesing,
  ``Simulation-optimization of digital twin,'' in {\em 2021 Winter Simulation
  Conference (WSC)}, pp.~1--10, IEEE, 2021.

\bibitem{rafiei2017new}
M.~H. Rafiei and H.~Adeli, ``A new neural dynamic classification algorithm,''
  {\em IEEE transactions on neural networks and learning systems}, vol.~28,
  no.~12, pp.~3074--3083, 2017.

\bibitem{xu2017implementation}
X.~Xu and W.~Chen, ``Implementation and performance optimization of dynamic
  random forest,'' in {\em 2017 International Conference on Cyber-Enabled
  Distributed Computing and Knowledge Discovery (CyberC)}, pp.~283--289, IEEE,
  2017.

\bibitem{aydemir2020digital}
H.~Aydemir, U.~Zengin, and U.~Durak, ``The digital twin paradigm for aircraft
  review and outlook,'' in {\em AIAA Scitech 2020 Forum}, p.~0553, 2020.

\bibitem{Khatamsaz2021}
D.~Khatamsaz and D.~L. Allaire, ``A comparison of reification and cokriging for
  sequential multi-information source fusion,'' in {\em AIAA Scitech 2021
  Forum}, p.~1477, 2021.

\bibitem{Chen2016}
S.~Chen, Z.~Jiang, S.~Yang, D.~W. Apley, and W.~Chen, ``Nonhierarchical
  multi-model fusion using spatial random processes,'' {\em International
  journal for numerical methods in engineering}, vol.~106, no.~7, pp.~503--526,
  2016.

\bibitem{Bostanabad2022}
N.~Oune, J.~T. Eweis-Labolle, and R.~Bostanabad, ``Data fusion with latent map
  gaussian processes,'' 2021.

\bibitem{gong2022data}
H.~Gong, S.~Cheng, Z.~Chen, and Q.~Li, ``Data-enabled physics-informed machine
  learning for reduced-order modeling digital twin: application to nuclear
  reactor physics,'' {\em Nuclear Science and Engineering}, pp.~1--26, 2022.

\bibitem{Chen2012}
S.~H. Chen and C.~A. Pollino, ``Good practice in bayesian network modelling,''
  {\em Environmental Modelling \& Software}, vol.~37, pp.~134--145, 2012.

\bibitem{chakraborty2021role}
S.~Chakraborty, S.~Adhikari, and R.~Ganguli, ``The role of surrogate models in
  the development of digital twins of dynamic systems,'' {\em Applied
  Mathematical Modelling}, vol.~90, pp.~662--681, 2021.

\bibitem{zotov2020towards}
E.~Zotov, A.~Tiwari, and V.~Kadirkamanathan, ``Towards a digital twin with
  generative adversarial network modelling of machining vibration,'' in {\em
  International Conference on Engineering Applications of Neural Networks},
  pp.~190--201, Springer, 2020.

\bibitem{xu2022design}
Z.~Xu, J.~Xu, Z.~Guo, H.~Wang, Z.~Sun, and X.~Mei, ``Design and optimization of
  a novel microchannel battery thermal management system based on digital
  twin,'' {\em Energies}, vol.~15, no.~4, p.~1421, 2022.

\bibitem{tuegel2012airframe}
E.~Tuegel, ``The airframe digital twin: some challenges to realization,'' in
  {\em 53rd AIAA/ASME/ASCE/AHS/ASC structures, structural dynamics and
  materials conference 20th AIAA/ASME/AHS adaptive structures conference 14th
  AIAA}, p.~1812, 2012.

\bibitem{Kononenko1989}
I.~Kononenko, ``Bayesian neural networks,'' {\em Biological Cybernetics},
  vol.~61, no.~5, pp.~361--370, 1989.

\bibitem{Sepahvand2010}
K.~Sepahvand, S.~Marburg, and H.-J. Hardtke, ``Uncertainty quantification in
  stochastic systems using polynomial chaos expansion,'' {\em International
  Journal of Applied Mechanics}, vol.~2, no.~02, pp.~305--353, 2010.

\bibitem{Koenker2001}
R.~Koenker and K.~F. Hallock, ``Quantile regression,'' {\em Journal of economic
  perspectives}, vol.~15, no.~4, pp.~143--156, 2001.

\bibitem{Zhou2019}
M.~Zhou, J.~Yan, and D.~Feng, ``Digital twin framework and its application to
  power grid online analysis,'' {\em CSEE Journal of Power and Energy Systems},
  vol.~5, no.~3, pp.~391--398, 2019.

\bibitem{syarif2012application}
I.~Syarif, E.~Zaluska, A.~Prugel-Bennett, and G.~Wills, ``Application of
  bagging, boosting and stacking to intrusion detection,'' in {\em
  International Workshop on Machine Learning and Data Mining in Pattern
  Recognition}, pp.~593--602, Springer, 2012.

\bibitem{Nae2019}
NAE, ``Engineering the future: 2019 annual report,'' Tech. Rep.~4, National
  Academy of Engineering, Washington, District of Columbia, 2019.

\end{thebibliography}
\end{document}